\documentclass[11pt, %showpacs, 
preprintnumbers,superscriptaddress,amsmath,amssymb,nofootinbib]{revtex4}
\usepackage{graphicx}% Include figure files
\usepackage{dcolumn}% Align table columns on decimal point
\usepackage{bm}% bold math
\usepackage{amssymb}
\usepackage{amsmath}
\usepackage{epsfig}    
\usepackage{color}
\usepackage{slashed}
\usepackage{hhline}
\def\be{\begin{equation}}
\def\ee{\end{equation}}
\newcommand{\bea}{\begin{eqnarray}}
\newcommand{\eea}{\end{eqnarray}}
\newcommand{\nn}{\nonumber}

\numberwithin{equation}{section}

\begin{document}

\title{A generalized one-loop neutrino mass model with charged particles}

\author{Kingman Cheung}
\email{cheung@phys.nthu.edu.tw}
\affiliation{Physics Division, National Center for Theoretical Sciences, 
Hsinchu, Taiwan 300}
\affiliation{Department of Physics, National Tsing Hua University, 
Hsinchu 300, Taiwan}
\affiliation{Division of Quantum Phases and Devices, School of Physics, 
Konkuk University, Seoul 143-701, Republic of Korea}

\author{Hiroshi Okada}
\email{macokada3hiroshi@cts.nthu.edu.tw}
\affiliation{Physics Division, National Center for Theoretical Sciences, 
Hsinchu, Taiwan 300}

%\pacs{}
\date{\today}

\begin{abstract}
We propose a radiative neutrino-mass model by introducing 3
generations of fermion pairs $E^{-(N+1)/2} E^{+(N+1)/2}$ and a couple
of multi-charged bosonic doublet fields $\Phi_{N/2}, \Phi_{N/2+1}$,
where $N=1,3,5,7,9$.  We show that the models can satisfy the neutrino
masses and oscillation data, and are consistent with lepton-flavor
violations, the muon anomalous magnetic moment, the oblique
parameters, and the beta function of the $U(1)_Y$ hypercharge gauge
coupling.  We also discuss the collider signals for various $N$,
namely, multi-charged leptons in the final state from the Drell-Yan
production of $E^{-(N+1)/2} E^{+(N+1)/2}$. In general, the larger
the $N$ the more charged leptons will appear in the final state.
\end{abstract}
\maketitle

\section{Introduction}
Nonzero neutrino mass is the most intriguing evidence for physics beyond the
standard model (SM). The scale of neutrino mass ($\sim 0.1$ eV) is at least
12 orders of magnitude smaller than the electroweak scale. In order to 
explain such a tiny neutrino mass, various mechanisms have been proposed
to explain the phenomena. Conventionally, the seesaw mechanism \cite{seesaw}
with a high seesaw scale ($\sim 10^{11-12}$ GeV) is one of the most
natural mechanisms to {generate} such a tiny mass.  
However, such a high seesaw scale
offers no immediate laboratory tests. Therefore, a number variety of seesaw
models appeared afterwards, e.g., inverse seesaw \cite{inverse}.  

Another category of models is the radiatively generated neutrino-mass models,
in which the smallness of neutrino mass is achieved by loop suppression.
A few of the earliest models are the Zee model \cite{zee}, Babu model 
\cite{babu}, {and Ma model \cite{ma}}.  In general, it requires new particles running in the loop(s) 
of the neutrino-mass generating diagrams. These new particles can be light
enough to be produced at colliders, thus offering immediate tests for the 
model.  They could also be relevant to explain other phenomena, 
such as dark matter, lepton-flavor violations, muon anomalous magnetic moment, 
etc. 

In a previous work, we proposed a simple extension of the SM 
by introducing 3 generations of doubly-charged fermion pairs $E^{--} E^{++}$ 
and three multi-charged bosonic fields $k^{++}, \Phi_{3/2}, \Phi_{5/2}$,
in which $\Phi$ are the $SU(2)$ doublet fields and $k^{++}$ is a singlet
field \cite{Cheung:2017kxb}. The model can explain the small neutrino masses and oscillations{, muon anomalous magnetic moment}, and is consistent with the lepton-flavor violations
and the oblique parameters.  

Here in this work we {generalize} the model to 
$E^{- \frac{N+1}{2} } E^{+ \frac{N+1}{2} }$ and $\Phi_{N/2}$, $\Phi_{N/2+1}$,
with $N=1,3,5,7,9$.  The previous work \cite{Cheung:2017kxb} 
corresponds to the case of $N=3$. The generalization is indeed nontrivial.
Especially, for the case of $N=1$ in which a $Z_2$ parity is required to
distinguish between 
the Higgs doublet $H$ and the doublet $\Phi_{1/2}$. The $Z_2$ assignment thus
gives rise to a lightest $Z_2$-odd particle, which is stable and can be a dark
matter candidate. We will explore the dark matter phenomenology of $N=1$ case. 
The other cases share some similar features as $N=3$, but they do have 
different features that deserve separate discussion.  
In general, a larger $N$ would rise  to a final state with more charged 
leptons for Drell-Yan production of $E^{+(N+1)/2} E^{-(N+1)/2}$. 

This paper is organized as follows.
In Sec.~II, we review the model and describe the constraints.
In Sec.~III, we describe the physics for each $N$.
In Sec. IV, we present the numerical analyses and valid parameter space
for each $N$.
We discuss the collider signals in Sec. V, and conclude in Sec.~VI.

\section{Model setup and Constraints with common part}

\begin{table}[t]
\begin{tabular}{|c|c|c|c||c|c|c|}
\hline\hline  
 &~$L_L$ ~&~$e_R$ ~&~$E_{L/R}$ ~& ~$H$~ & ~$\Phi_{{N}/2}$~  & ~$\Phi'_{{N'}/2}$ \\\hline 
$SU(2)_L$ & $\bm{2}$& $\bm{1}$& $\bm{1}$   & $\bm{2}$ & $\bm{2}$  & $\bm{2}$  \\\hline 
$U(1)_Y$   & $-\frac12$ & $-1$ & $-\frac{N+1}2$ & $\frac12$ & $\frac{N}2$ & $\frac{N'}2$    \\\hline
\end{tabular}
\caption{Charge assignments of new fields under $SU(2)_L\times U(1)_Y$ with $1 \le N$ and $N'\equiv N+2$ with odd number, where all the new fields are color singlet.}
\label{tab:1}
\end{table}

In our set up of the model, we introduce three families of
doubly-charged fermions $E$~\footnote{In order to minimally reproduce
the neutrino oscillation data, two families of $E$ are enough. In
this case, a massless neutrino is induced.}, and two types of new
bosons $\Phi_{N/2}$ and $\Phi_{{N'}/2}$ with $N'\equiv N+2$, as shown
in Table~\ref{tab:1}.  
Notice here that one has to impose an
additional symmetry such as $Z_2$ to discriminate between $\Phi_{N/2}$
and $H$ only in case of $N=1$, as we will see later.  The
renormalizable Lagrangian in the lepton sector and the Higgs potential
are given by
\begin{align}
-\mathcal{L}_{Y}
&= (y_\ell)_{ii} \bar L_i H P_R e_{i}
+ f_{ia} \bar L_i\Phi_{N/2}   P_R E_a  %\nn\\&
+ g_{ia} \bar L_i  \cdot \Phi'^*_{N'/2} P_R E^c_a + M_{E_a} \bar E_a E_a + {\rm h.c.}, \nn\\
%%%
V&=
\mu_H^2 |H|^2 +\mu_\Phi^2 |\Phi_{N/2}|^2 + \mu_{\Phi'}^2 |\Phi'^2_{N'/2}|\nn\\
&+\left[\lambda_0 (H^T\cdot \Phi_{N/2}) (H^T\cdot \Phi'^*_{N'/2}) +{\rm c.c.}\right]
+\left[\lambda_0' (\Phi'^\dag_{N'/2} \Phi_{N/2})_{\bf 3} (H^T H)_{\bf 3} +{\rm c.c.}\right]\nn\\
& +\lambda_H |H|^4 + \lambda_\Phi |\Phi_{N/2}|^4+ \lambda_{\Phi'} |\Phi'_{N'/2}|^4 + \lambda_{\Phi\Phi'} |\Phi_{N/2}|^2 |\Phi'_{N'/2}|^2
+ {\rm h.c.},\label{Eq:lag-yukawa} 
%%%%%%%%%%%%%%
\end{align}
where $(i,a)=1-3$ are generation indices and the multiplication
symbol $\cdot$ represents $i\sigma_2$ with $\sigma_2$ being the second Pauli matrix.
The first term in the Yukawa Lagrangian, which is assumed to be diagonal for convenience, 
provides the masses for the charged leptons {$(m_{\ell_i}\equiv y_{\ell_{ii}}v/\sqrt2$)} by developing a nonzero 
vacuum expectation value (VEV) of $H$, which is symbolized 
by $\langle H\rangle\equiv v/\sqrt2$.
Also, we work in the basis where all the coefficients are real and 
positive for simplicity hereafter. 
We can parameterize the scalar fields as
\begin{align}
&\Phi_{N/2} =\left[
\begin{array}{c}
\phi^{\frac{N+1}{2}}\\
\phi^{\frac{N-1}{2}}
\end{array}\right],
\quad 
%%%
\Phi'_{N'/2} =\left[
\begin{array}{c}
\phi'^{\frac{N+3}{2}} \\
\phi'^{\frac{N+1}{2}}
\end{array}\right],
\label{eq:component}
\end{align}
where the superscript for each component represents 
the electric charges.
%%%
Due to the $\lambda_0^{(')}$ term in Eq.~(\ref{Eq:lag-yukawa}), the 
two $\frac{N+1}{2}$-charged bosons in basis of 
$(\phi^{\frac{N+1}{2}},\phi'^{\frac{N+1}{2}})$ mix with each other. Their 
mixing matrix and mass eigenstates are defined as 
\begin{align}
%%%
&\left[\begin{array}{c} \phi^{\frac{N+1}{2}} \\ \phi'^{\frac{N+1}{2}}\end{array}\right] = 
\sum_{a=1-2}O_{ia} H_{a} %^{\frac{N+1}{2}}
,\quad
%%%
O\equiv 
%%%
\left[\begin{array}{cc} 
 c_{\theta}  & -s_{\theta} \\
  s_{\theta}  & c_{\theta}   
  \end{array}\right].
\end{align}
Therefore  one can redefine these bosons as the mass eigenstates as follows:
\begin{align}
%%%
\phi^{\frac{N+1}{2}} = c_{\theta} H_1%^{\frac{N+1}{2}} 
-  s_{\theta} H_2%^{\frac{N+1}{2}}
,\quad
\phi'^{\frac{N+1}{2}} = s_{\theta} H_1%^{\frac{N+1}{2}} 
+  c_{\theta} H_2, %^{\frac{N+1}{2}}.
 \label{eq:basis}
\end{align}
where  we have used the short-hand notation: 
$H_i\equiv H_i^{\frac{N+1}{2}}$ ($i=1,2$), and their masses {to} 
be $m_{H_i^{\frac{N+1}{2}}}\equiv m_{H_i}$.

%%%%%%%%%%%%%%%%%%
\begin{figure}[tb]
\begin{center}
\includegraphics[width=80mm]{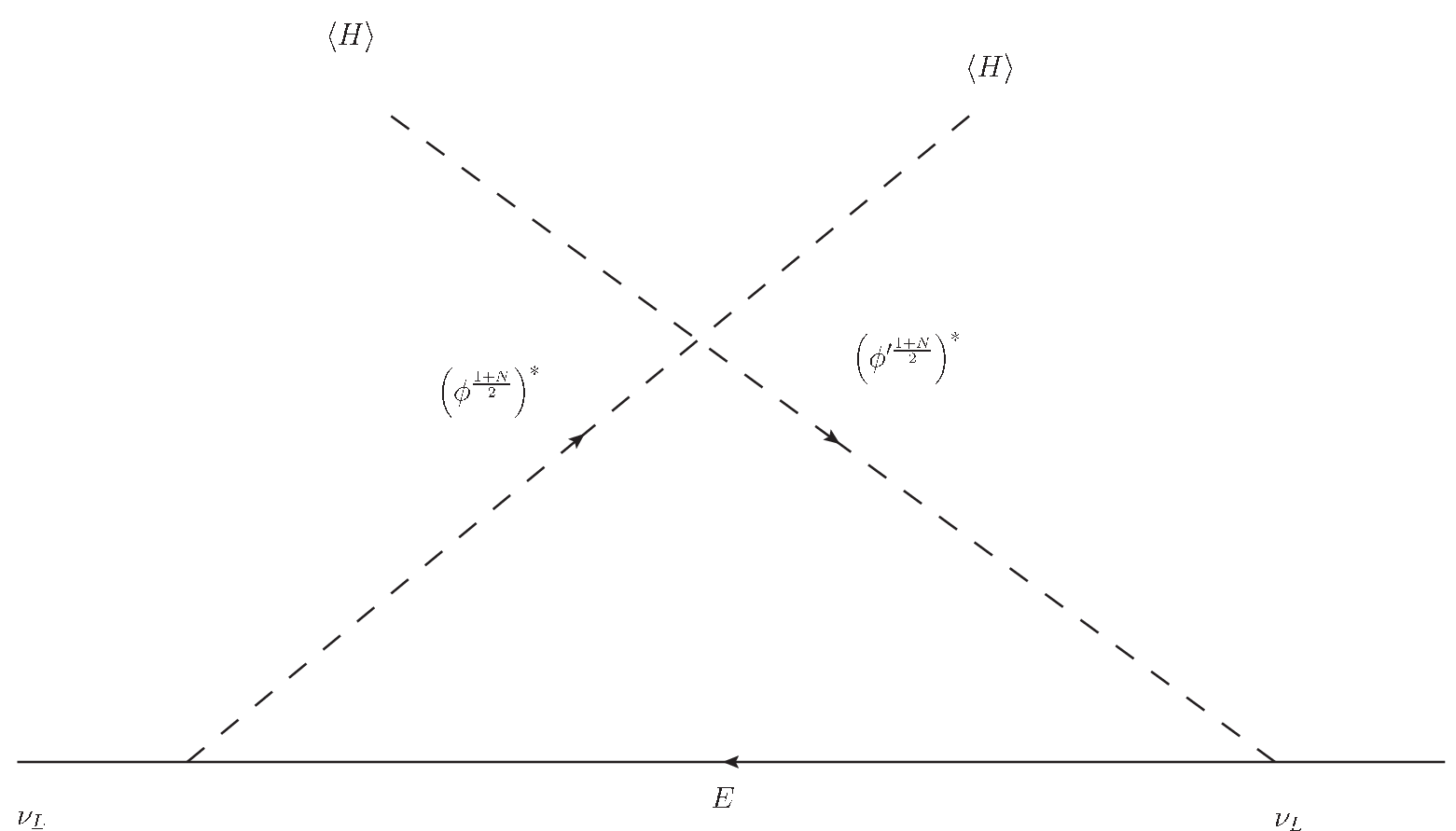}
\caption{ One-loop diagrams for generating the neutrino mass matrix.}
\label{fig:neutrino}
\end{center}
\end{figure}
%%%%%%%%%%%%%%%%%%%

\subsection{Neutrino mixing}
The active neutrino mass matrix $M_\nu$
is given at one-loop level via doubly-charged particles in Fig.~\ref{fig:neutrino}, and its formula is given by~\cite{Cheung:2017kxb}
%\begin{widetext}
\begin{align}
- (M_{\nu})_{ij}
&=\frac{2 s_{2\theta}}{(4\pi)^2}\sum_{a=1}^3
{ f_{ia} M_a g_{aj}^T }%{(4\pi)^2}
 F_I(M_a,H_1, H_2) + (f\leftrightarrow g)  \equiv  f_{ia} R_a g_{aj}^T + g_{ia} R_a f_{aj}^T
,\\
%%%
F_I(m_a,m_b,m_c)
&=\frac{m_a^2 m_b^2\ln\left(\frac{m_a}{m_b}\right)+m_a^2 m_c^2\ln\left(\frac{m_a}{m_c}\right)+m_b^2 m_c^2\ln\left(\frac{m_b}{m_c}\right)}
{(m_a^2-m_b^2)(m_a^2-m_c^2) }.
\end{align}%\end{widetext}
The mass matrix $M_\nu$ is diagonalized by the neutrino mixing 
matrix $V_{\rm MNS}$ as 
$M_\nu  = V_{\rm MNS} D_\nu V_{\rm MNS}^T$ with $D_\nu\equiv 
(m_{\nu_1},m_{\nu_2},m_{\nu_3})$. 
Then one can parameterize Yukawa coupling in terms of arbitrary antisymmetric matrix with complex values $A$; $(A+A^T=0)$, as 
follows~\cite{Okada:2015vwh}
\begin{align}
& f=-\frac12 [V_{\rm MNS} D_\nu V_{\rm MNS}^T+A] (g^T)^{-1} R^{-1},
\quad
 g=-\frac12 [V_{\rm MNS} D_\nu V_{\rm MNS}^T+A]^T (f^T)^{-1} R^{-1}.\label{eq:g-to-f}
\end{align}
In the numerical analysis, we shall use {the latter} relation for
convenience, and use the
data in the global analysis~\cite{Forero:2014bxa}.

{ 
It is worthwhile to describe the differences between the Ma model and our
model.  Once we assign the lepton number $+2$ for $\Phi'_{N'/2}$ and
zero for other bosons, the lepton number is explicitly broken at
the $\lambda_{0}$ and $\lambda_{0}^{'}$ terms by 2 units. 
On the other hand, if all the bosons were assigned zero charges
under the lepton number, then the breaking of lepton number would
occur at the Yukawa coupling term $g$. 
Here, we suppose that the extra
vector fermions $E$ has $-1$ charge under the lepton number, which is
the same as the SM charged leptons.  
The major difference between our model and the Ma model is that
the lepton number in the Ma model is violated 
via the mass term of the right-handed Majorana neutrinos
while in our model it is violated via the boson mass or the Yukawa term.
}

\subsection{Lepton flavor violations and muon anomalous magnetic moment}
\label{lfv-lu}
The Yukawa terms of ($f,g$) give rise to $\ell_i\to\ell_j\gamma$ processes 
at one-loop level.
%but also the  muon anomalous magnetic moment.
%$Considering its measured muon $g-2$ is positively obtained, $f$ should be large.
%However since $\kappa$ does not contribute to to the neutrino sector and ,
 %
The branching ratio is given by
\begin{align}
B(\ell_i\to\ell_j \gamma)
\approx 
\frac{48\pi^3 \alpha_{\rm em}}{{G_{\rm F}^2} } C_{ij} |{\cal M}_{ij}|^2,
\end{align}
where $G_{\rm F}\approx1.166\times 10^{-5}$ GeV$^{-2}$ is the Fermi constant, 
$\alpha_{\rm em}(m_Z)\approx {1/128.9}$ is the 
fine-structure constant~\cite{Agashe:2014kda}, 
$C_{21}\approx1$, $C_{31}\approx0.1784$, and $C_{32}\approx0.1736$.
${\cal M}(= {\cal M}_f+{\cal M}_g)$ is formulated as 
\begin{align}
&({\cal M}_f)_{ij}
\approx
-\sum_{a=1-3}\frac{f_{ja} f^\dag_{ai} }{(4\pi)^2} 
\left[ \frac{N-1}{2}  {F_{lfv}(M_{E_a} , m_{\phi}) +\frac{N+1}{2} F_{lfv}}(m_{\phi},M_{E_a})\right], \label{eq:lfv-f}\\
%%%
&({\cal M}_g)_{ij}
\approx
\sum_{a=1-3}\frac{g_{ja} g^\dag_{ai} }{(4\pi)^2}
\left[\frac{N+3}{2} {F_{lfv}(M_{E_a}, m_{\phi'}) +\frac{N+1}{2} F_{lfv}}(m_{\phi'} , M_{E_a})\right], \label{eq:lfv-g}\\
%%%
&F_{lvs}(m_1,m_2)\approx \frac{2 m_1^6+3m_1^4m_2^2-6m_1^2m_2^4+m_2^6+12m_1^4 m_2^2\ln\left[\frac{m_2}{m_1}\right]}
{12(m_1^2-m_2^2)^4},\label{eq:lfv-lp}
%%%
\end{align} 
where we have simplified the notation 
$\phi \equiv \phi^{\frac{N-1}{2}}$, and $\phi'\equiv \phi'^{\frac{N+3}{2}}${, and assumed $m_{\ell_\tau}<<M_{E_a},m_\phi,m_{\phi'}$ in the equations above.}\footnote{{It implies that the mass difference between $m_1$ and $m_2$ should be greater than the order of $m_{\ell_\tau}$.}}
%%%
The current experimental upper bounds are given 
by~\cite{TheMEG:2016wtm, Aubert:2009ag}
  \begin{align}
  B(\mu\rightarrow e\gamma) &\leq4.2\times10^{-13},\quad 
  B(\tau\rightarrow \mu\gamma)\leq4.4\times10^{-8}, \quad  
  B(\tau\rightarrow e\gamma) \leq3.3\times10^{-8}~.
 \label{expLFV}
 \end{align}

%%%%%%%%%%%%%%%%%
{\it The muon anomalous magnetic moment ($\Delta a_\mu$)}: 
We can also estimate the muon anomalous magnetic moment through ${\cal M}$, which is given by 
\begin{align}
\Delta a_\mu\approx -m_\mu^2 {\cal M}_{22}.\label{eq:damu}
\end{align}
The $3.3\sigma$ deviation from the SM prediction is  
$\Delta a_\mu=(26.1\pm8)\times10^{-10}$~\cite{Hagiwara:2011af} with 
a positive value. 
Obviously, $f$ contributes to the $\Delta a_\mu$ positively, 
while $g$ does negatively.
To achieve the agreement with experimental result, one has to enhance 
the contributions from $f$ compared to those from $g$.

%%%%%%%%%%%%%%%%%%
\begin{figure}[tb]
\begin{center}
\includegraphics[width=100mm]{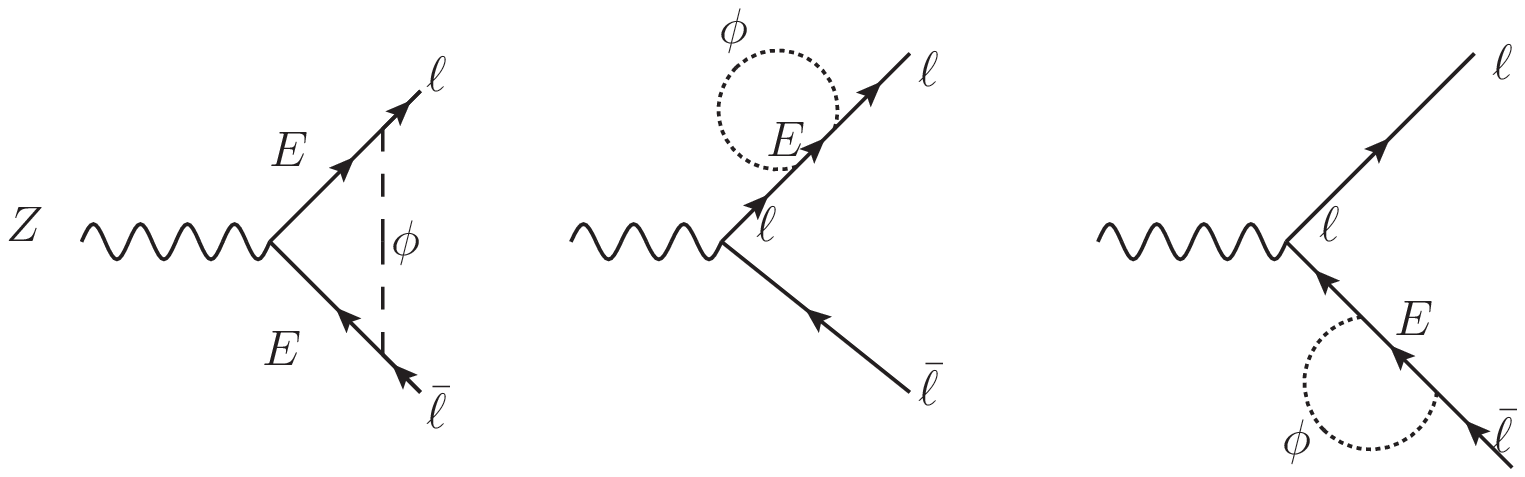}
\includegraphics[width=100mm]{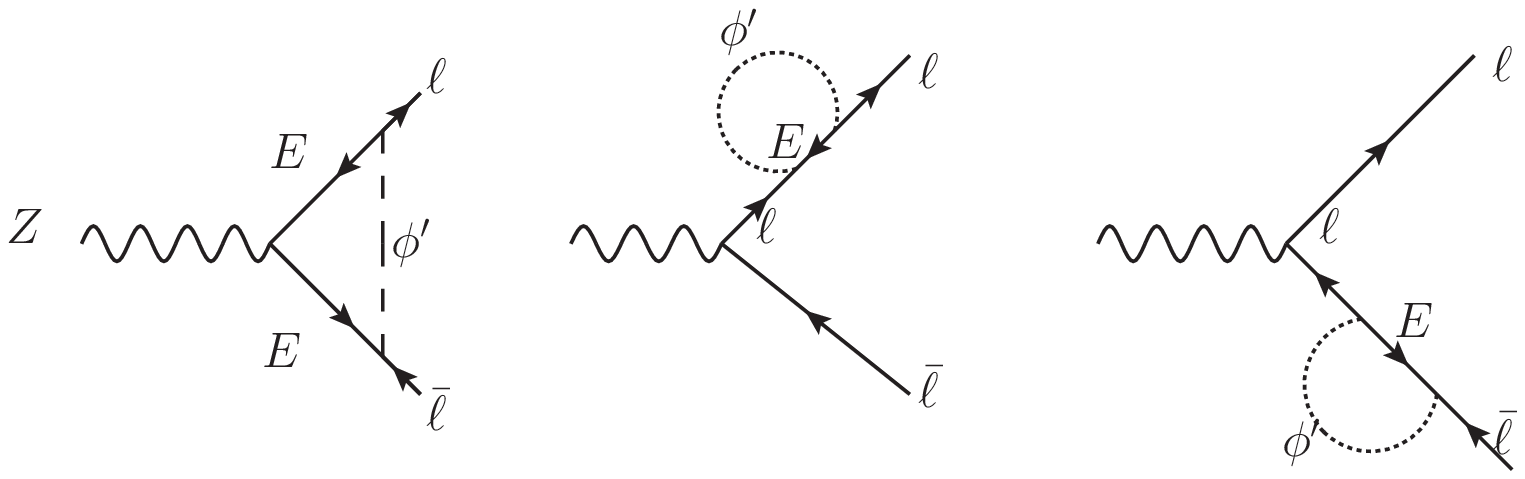}
\caption{Feynman diagrams for $Z\to \ell_i\bar\ell_j$, where upper diagrams represent the contributions for $f$, while the down ones for $g$.}
\label{fig:zto2ell}
\end{center}
\end{figure}
%%%%%%%%%%%%%%%%%%%

%%%%%%%%%%%%%%%%%%%%%%%%%%%%%%%%%%%%%%%%%%%%%%%%%%
\subsection{Flavor-Changing Leptonic $Z$ Boson Decays}\label{subsec:Zll}
%%%%%%%%%%%%%%%%%%%%%%%%%%%%%%%%%%%%%%%%%%%%%%%%%%
%%%
In this subsection, we consider the $Z$ boson decay into two charged
leptons with different flavors through the Yukawa terms $f$ and
$g$ at one-loop level~\cite{Chiang:2017tai}.
%~\footnote{$g$ also generally contribute to the $Z$ boson decay at one loop level, and its obtained by changing $\phi\to\phi'$ and $f\to g$. However since we expect $g$ to be tiny to suppress the negative contribution to $(g-2)_\mu$, we do not consider hereafter.}
Since some components of $f$ and $g$ are expected to be large so as to obtain 
a sizable $\Delta a_\mu$, the experimental bounds on lepton-flavor violating
$Z$ boson decays could be of concern at one loop level. 
First of all, the relevant Lagrangian is given by
\begin{align}
{\cal L}_{kin.}\sim\frac{g_2}{c_w} \left[\bar\ell\gamma^\mu \left(-\frac12+s_w^2\right)\ell+\frac{N+1}{2}s_W^2 \bar E\gamma^\mu E \right] Z_\mu
+ {\left[ 
f_{ia} \bar\ell_i P_R E_a \phi - g_{ia} \bar\ell_i P_R E^c_a \phi'^*+{\rm h.c.}
\right] }
,
\end{align}
where $s(c)_W\equiv\sin(\cos)\theta_W\sim0.23$ 
stands for the sine (cosine) of 
the Weinberg angle. 
%The amplitudes of such decay modes involve the Yukawa couplings $f_{ia}$, some of which can be of ${\cal O}(1)$ in order to achieve a sizable contribution to the muon $g-2$. 
%%%
Combining all the diagrams in Fig.~\ref{fig:zto2ell},
the ultraviolet divergence cancels out and only the finite part 
remains~\cite{Chiang:2017tai}.
The resulting form is given by
\begin{align}
%{\color{blue}
&\text{BR}(Z\to\ell^-_i\ell^+_j)
=
\frac{G_F}{3\sqrt2 \pi} \frac{m_Z^3 s_w^4(N+1)^2}{4(16\pi^2)^2 \Gamma_{Z}^{\rm tot}} \nn\\
%\left(s_W^2 -\frac12\right)^2 \nn\\& \qquad\qquad \times
&\hspace{1cm}\times
\left|
\sum_{a=1}^{3} f_{ia} f^*_{ja} \left[ F_2(E_a,\phi)+F_3(E_a,\phi) \right]
+
\sum_{b=1}^{3} g_{ib} g^*_{jb} \left[ F_2(E_b,\phi')+F_3(E_b,\phi')\right] 
\right|^2, \label{eq:Zll}
%}
\end{align}
where $i\neq j$,
\begin{align*}
F_2(a,b) &=\int_0^1dx(1-x)\ln\left[ (1-x)m_a^2 + x m_b^2 \right] ~,
\\
F_3(a,b) &=\int_0^1dx\int_0^{1-x}dy\frac{(xy-1)m_Z^2+(m_a^2-m_b^2)(1-x-y)-\Delta\ln\Delta}{\Delta} ~,
\end{align*} 
with $\Delta\equiv -xy m_Z^2+(x+y)(m_a^2-m_b^2)+m_b^2$ and the total $Z$ decay width $\Gamma_{Z}^{\rm tot} = 2.4952 \pm 0.0023$~GeV~\cite{Patrignani:2016xqp}.
From Eqs.~(\ref{eq:lfv-f}), (\ref{eq:lfv-g}) and (\ref{eq:Zll}), one finds 
that the same combinations of Yukawa couplings
$f_{ia} {f^*_{ja}}$ and $g_{ia} {g^*_{ja}}$ appear 
in $\ell_i\to\ell_j\gamma$ and $Z\to \ell_i \ell_j $.
%However, a crucial difference is their decoupling properties. 
%{The former modes would be suppressed as the particles in the loops become heavy while the latter can grow logarithmically.  This difference may stem from the different structures in the form factors: the former of the dipole type and the latter of the vector one.  A similar nondecoupling behavior of the LFV $Z$ decays can be found in Ref.~\cite{Illana:2000ic}, 
Note that the predictions for $\text{BR}(Z\to\mu\tau)$ can be rather 
large and tested in future experiments such as 
Giga-Z~\cite{AguilarSaavedra:2001rg}, due to different properties
of loop functions~\cite{Chiang:2017tai}.
%%%
The current bounds on the lepton-flavor-changing $Z$ boson decay 
branching ratios at 95 \% CL are given by \cite{Patrignani:2016xqp}:
\begin{align}
%\begin{split}
  {\rm BR}(Z\to e^\pm\mu^\mp) < 1.7\times10^{-6} ~,\
  {\rm BR}(Z\to e^\pm\tau^\mp) < 9.8\times10^{-6} ~,\
  {\rm BR}(Z\to \mu^\pm\tau^\mp) < 1.2\times10^{-5} ~.\label{eq:zmt-exp}
%\end{split}
\end{align}
We include these constraints in the global analysis.
%We have scanned the parameter space and found that all these constraints are less stringent than those from the LFV processes,
%{as well as the flavor-conserving processes ${\rm BR}(Z\to \ell^\pm\ell^\mp)$ ($\ell=e,\mu,\tau$). }

\subsection{ Oblique parameters} 
In order to estimate the testability via collider physics, we have to
consider the oblique parameters that restrict the mass hierarchy
between each of component $\Phi_{ N/2}$ and $\Phi_{ {N'}/2}$.

Here we focus  on the new physics contributions to the $S$ and $T$ parameters 
in the case of $\Delta U=0$.
Then  
$\Delta S$ and $\Delta T$ are defined as
\begin{align}
\Delta S&={16\pi} \frac{d}{dq^2}[\Pi_{33}(q^2)-\Pi_{3Q}(q^2)]|_{q^2\to0},\quad
\Delta T=\frac{16\pi}{s_{W}^2 m_Z^2}[\Pi_{\pm}(0)-\Pi_{33}(0)],
\end{align}
where $s_{W}^2\approx0.23$ is the Weinberg angle and $m_Z$ is the $Z$ 
boson mass. 
The loop factors $\Pi_{33,3Q,\pm}(q^2)$ are calculated from the one-loop 
vacuum-polarization diagrams for $Z$ and $W^\pm$ bosons, which are respectively given by~\cite{Cheung:2016fjo, Cheung:2017lpv}
\begin{align}
\Pi_{33}(q^2)&=\frac1{2(4\pi)^2}
\left[
G(\phi,\phi)
+G({H_\alpha},{H_\alpha})
+{G( \phi',\phi')}
%%%
-{H({\phi})}-
{H({H_\alpha})}
-{H({\phi'})}\right], \label{eq:pi33}\\
%%% %%%%%% %%%
\Pi_{3Q}(q^2)&=
\frac1{(4\pi)^2}
\left[ -m G(q^2,\phi,\phi) +(m+2)G(q^2, \phi',\phi') + m H(\phi) - (m+2)H(\phi').\right.\nn\\
&\left.
+(m+1)\left[ (c_\theta^2-s^2_\theta)(G(q^2,H_1,H_1)-G(q^2,H_2,H_2)) -2c_\theta s_\theta (G(q^2,H_1,H_2) + G(q^2,H_2,H_1)) \right]
\right.\nn\\
%%%
&\left. 
-(m+1)(c_\theta^2-s^2_\theta) \left(H(H_1)-H(H_2) \right)
\right],
\\
%%% %%%
\Pi_{\pm}(q^2)&=
\frac1{(4\pi)^2}
\left[
c^2_\theta G(q^2,\phi,H_1) + s^2_\theta G(q^2,\phi,H_2)
+s^2_\theta G(q^2,H_1,\phi') + c^2_\theta  G(H_2,\phi') \right.\nn\\
%%%
&\left.
-\frac12\left[
H(\phi)+H(\phi') +H(H_1) + H(H_2)\right]
\right],\\
%%%%%%%%%%%%
G(q^2,m1,m2)&\equiv\int[dX]_2[-q^2x(1-x)+xm_1^2+y m_2^2]
\left(\Upsilon+1-\ln\left[-\frac{q^2}{m_1^2}x(1-x)+x+y\frac{m_2^2}{m_1^2}\right]\right),
%%%
\end{align}
where $H(m)\equiv m^2[\Upsilon+1]$~\footnote{Notice here that $H(m)$ does not depend on the referenced energy $q^2$ that arises from contact interacting loop functions.}, $\int[dX]_2\equiv \int_0^1dxdy\delta(1-x-y)$, $\Upsilon\equiv \frac1\epsilon-\gamma-\ln(4\pi)$,
%$G(a,b)$ and $H(c)$ are defined in Appendix of ref.~\cite{Cheung:2016fjo},
$m\equiv \frac{N-1}{2}$ is electric charge, and $\alpha(=1,2)$ should be summed up.
Fixing $\Delta U=0$, the experimental bounds on $\Delta S$ and $\Delta T$ 
are given by \cite{Baak:2012kk}
\begin{align}
\Delta S = (0.05 \pm 0.09), \quad 
 \Delta T = (0.08 \pm 0.07),
 \end{align}
 with a correlation coefficient of $+0.91$. The $\Delta \chi^2$ can be 
calculated as~\cite{Dawson:2009yx}
 \begin{align}
 \Delta \chi^2=\sum_{(i,j)=1,2}(\Delta S-0.05,\Delta T-0.08)
%%%
\left[
\begin{array}{cc}
718.19 & -840.28 \\
 -840.28 & 1187.2 \\
\end{array}\right]
%%% 
\left[
\begin{array}{c}
\Delta S-0.05 \\ \Delta T-0.08 
\end{array}\right],
 \end{align}
 %%% 
and we impose the 99\% confidence level limit that corresponds to 
$\Delta \chi^2=9.210$ in our numerical analysis.

\subsection{Beta function of $g_Y$}
\label{beta-func}
\begin{figure}[tb]
\begin{center}
\includegraphics[width=13cm]{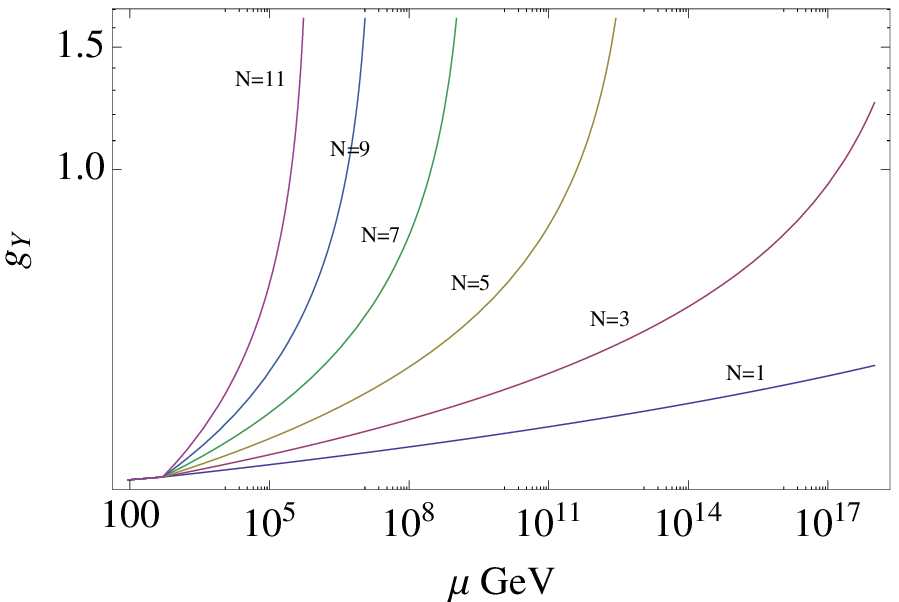}
\caption{The running of $g_Y$ in terms of a reference energy of $\mu$, depending on each of $N=1,3,5,7,9,11$.}
\label{fig:rge}
\end{center}\end{figure}

%%%
Here we estimate the effective energy scale by evaluating the Landau
pole for $g_Y$ in the presence of new exotic fields with nonzero
multiple hypercharges.  Each contribution of the new beta function 
of $g_Y$ from one 
$SU(2)_L$ doublet fermion or boson with $\pm N/2$ hypercharge is given
by~\cite{Ko:2016sxg}
\begin{align}
\Delta_s b^f_Y={\frac{3}{5}\times}\frac{4}{3}\times\left(\frac{N}2\right)^2 \ ,\quad \Delta_s b^b_Y={\frac{3}{5}\times}\frac{2}{3}\times\left(\frac{N}2\right)^2 \ ,
\end{align}
where the superscript of $\Delta b$ represents the fermion (f) or the 
boson (b), respectively.
Similarly, the contribution to the beta function from 
one $SU(2)_L$ singlet boson with $\pm N/2$ hypercharge($=$electric charge)
is given by 
\begin{align}
\Delta_d b^f_Y={\frac{3}{5}\times}\frac{2}{3}\times\left(\frac{N}2\right)^2 \ ,\quad \Delta_d b^b_Y={\frac{3}{5}\times}\frac{1}{3}\times\left(\frac{N}2\right)^2  ,
\end{align}
{where $3/5$ is the rescaled coefficient.}
%%%
Here let us include a doubly-charged gauge-singlet boson $k^{\pm\pm}$ 
in order to make appropriate decays into the SM fields as we will discuss 
later.~\footnote{$k^{++}$ is valid for $N=3,5,7,9$ to make appropriate 
decays, where it is not needed for $N=1$ due to existence of the 
DM candidate but an additional symmetry such as $Z_2$ is needed.}
Then one finds the energy evolution of the gauge coupling 
$g_Y$ as~\cite{Kanemura:2015bli}
\begin{align}
\frac{1}{g^2_Y(\mu)}&=\frac1{g_Y^2(m_{in.})}-\frac{b^{SM}_Y}{(4\pi)^2}\ln\left[\frac{\mu^2}{m_{in.}^2}\right]\nn\\
&
-\theta(\mu-m_{thres._f}) \frac{\Delta b^f_Y}{(4\pi)^2}\ln\left[\frac{\mu^2}{m_{thres._f}^2}\right]
-\theta(\mu-m_{thres._b}) \frac{\Delta b^b_Y}{(4\pi)^2}\ln\left[\frac{\mu^2}{m_{thres._b}^2}\right],\label{eq:rge_gy}
\end{align}
where $\mu$ is a reference energy scale, $b^{SM}_Y=41/6$, $\Delta b^f_Y=3
\Delta_s b^f_Y=(N+1)^2/2$, $\Delta b^b_Y=\Delta_d
b^b_Y(\Phi_{N/2})+\Delta_d b^b_Y(\Phi'_{N'/2})+\Delta_s
b^b_Y(k^{++})=(N^2+2N+6)/3$, and we assume that
$m_{in.}(=m_Z)<m_{thres._f}=m_{thres._b}=$500 GeV, where the
threshold masses for exotic fermions and bosons are $m_{thres._f}$ and
$m_{thres._b}$, respectively.
%%%
%Here we include contributions from three exotic singlet fermions $E$, new doublet scalar $\Phi_{N/2}(\Phi'_{N'/2})$, and a doubly charged gauge singlet scalar $k^{\pm\pm}$ that plays a crucial role in making appropriate decays into the SM fields as we will discuss later.
The resulting running of $g_Y(\mu)$ versus the scale $\mu$ 
is shown in Fig.~\ref{fig:rge} for 
each of $N=1,3,5,7,9,11$, where 
we analyze $N$ up to 9 in our global analyses.
\footnote{When one considers the case of $11\le N$, one has to
  introduce more particles in order to make appropriate decays of exotic 
particles into the SM particles. 
For example, the minimum extension in the case of $N=11$ is to
  add a gauge-singlet quartic-charged boson $h^{\pm\pm\pm\pm}$.
  In this case, the valid scale (=at the Landau pole) 
  decreases down to $10^4$ GeV. }
%%%
In the cases of $N=1,3$, perturbativity allows
 the cutoff scale up to 
Planck mass.  However, the case of $N=5,7,9,11$ is valid up to 
around $3\times10^{11}$ GeV, $3\times10^8$ GeV, $3\times10^5$ GeV, 
and $10^5$ GeV, respectively.
Notice here that $g_2$ running is almost the same the one of SM.

\section{Physics of Each $N$}
Here we investigate each of $N(1,3,5,7,9)$ below, where we introduce a
gauge-singlet doubly-charged boson $k^{++}$ for $3\le N$.
{\it We assume that the mixing between $k^{++}$ and a
component of doubly-charged boson with isospin doublet $\Phi_{N/2}$
and/or $\Phi'_{N'/2}$ are small enough to be neglected for the
neutrino oscillations, LFVs, and $\Delta a_\mu$, although the mixing
could play an important role in appropriately making exotic fields
decay into the SM fermions.}
%~\footnote{As an example to consider this mixing, see Ref.~\cite{Cheung:2017kxb}.} }

\subsection{$N=1$}
The case of $N=1$ is different from other $N$'s, because one cannot distinguish
between the SM Higgs field $H$ and $\Phi_{N/2}$ unless they are distinguished 
by some additional symmetries.  This category of models has been analyzed  
in Ref.~\cite{Aoki:2011yk}, which we follow and impose
a $Z_2$ symmetry on the new exotic fields $E$, $\Phi_{1/2}$, and
$\Phi'_{3/2}$. 
The corresponding Lagrangian and potential retain the general forms 
in Eqs.~(\ref{Eq:lag-yukawa}).  
Here we focus on the analysis of the DM candidate and explore the 
allowed region combining all the predictions and constraints as 
discussed before.  
Since the relic density from the kinetic term had been analyzed extensively in
Ref.~\cite{Hambye:2009pw}, we neglect the contributions from
the potential terms with small couplings in order to satisfy the
constraints of direct detection experiments such as
LUX~\cite{Akerib:2016vxi}, XENON1T~\cite{Aprile:2017iyp}, and
PandaX-II~\cite{Cui:2017nnn}. 
Therefore, we concentrate on the Yukawa coupling term $f$ with the mass 
range $m_Z/2\lesssim M_X\lesssim
m_W$.~\footnote{The lower bound comes from forbidding $Z\to 2X$ decay
  which is strongly constrained by precision experiments, while the
  upper bound comes from the assumption that the kinetic terms does not
  contribute significantly to the relic density of DM.}
%%%
Moreover, we have to give a larger mass difference between CP-even and
CP-odd bosons of $\phi^0\sim \phi_R+i\phi_I$, (either of which can be the DM),
in order to evade large cross sections via interaction of
$Z-\phi_R-\phi_I$, otherwise the coannihilation that gives a very small relic
density.~\footnote{The typical mass difference is about 10\%$\sim$20
  \% of the DM mass~\cite{Griest:1990kh}, and the mass difference can
  arise from the additional term $(H^\dag \Phi_{1/2})^2+{\rm
    c.c.}$.}
%%%
The relevant interaction of the DM $\phi_R=X$ is then given by
%~\footnote{Notice here that we have an additional term $(H^\dag \Phi_{1/2})^2+{\rm c.c.}$ that induces mass difference between real and imaginary part of $\phi^0$ just like Ma model~\cite{Ma:2006km}. However we assume to be negligible for simplicity.}
\begin{align}
-{\cal L}=\frac{f_{ia}}{\sqrt2} \bar\ell_i P_R E_a X +{\rm h.c.}.
\end{align}
Then the dominant contribution is found to be the effective 
d-wave coefficient after expanding the cross section in terms of 
the relative velocity. Its form is given by~\cite{Nomura:2017psk}
\begin{align}
d_{eff}(2X\to f_i \bar f_j)\approx 
\frac{M_X^6}{240 \pi}\left|\sum_{a=1}^3 \frac{f_{ia}f^\dag_{aj}}{(M_X^2+M^2_{E_a})^2}\right|^2,
%\frac{M_X^6}{240 \pi}\sum_{a=1}^3  \frac{ |f_{ia}f^\dag_{aj}|^2}{(M_X^2+M^2_{E_a})^4},
\end{align}
where $(f_i, f_j)=(e,\mu,\tau)$ run over all the charged-lepton 
mass eigenstates, and we have taken the charged-lepton masses in the final 
state to be zero. The resulting relic density is given by
\begin{align}
\Omega h^2\approx \frac{6.852\times 10^{-9}}{{\rm GeV^2}}\frac{1}{d_{eff}}.
\end{align}
where we have used several physical parameters given 
in Ref.~\cite{Nomura:2017psk}.  The measured relic density is 
about 0.12~\cite{pdg}, and we apply the constraint as 
$0.11\lesssim \Omega h^2\lesssim 0.13$ in our numerical analysis.

Since this mode is d-wave suppressed, so a rather large $f$ is required,
which is in favor of generating a sizable $\Delta a_\mu$.
%%%
%In Fig.~\ref{xxx}, we show the correlation between relic density and $(g-2)_\mu$.
%%%%%%%%
Notice here that we have to replace $\phi$ by  $\phi_R/\sqrt2$ and 
$\phi_I/\sqrt2$, 
due to the mass difference between  $\phi_R$ and $\phi_I$. 
Hence all the general forms, except for the neutrino mass matrix,
should be modified according to the replacement. 
For example, $F_{lfv}(M_{E_a},m_\phi)\to [F_{lfv}(M_{E_a},m_{\phi_R}) + 
F_{lfv}(M_{E_a},m_{\phi_I}) ]/2$ in Eq.~(\ref{eq:lfv-f}), and 
(\ref{eq:lfv-g}), $F_{2,3}(E_a,\phi)\to[F_{2,3}(E_a,\phi_R)+F_{2,3}
(E_a,\phi_I)]/2$ in  Eq.~(\ref{eq:Zll}), 
$G(\phi,\phi)\to[G(\phi_R,\phi_R)+G(\phi_I,\phi_I)]/2$ 
and $H(\phi)\to[H(\phi_R)+H(\phi_I)]/2$ in Eq.~(\ref{eq:pi33}).

\subsection{$N=3$}
The $N=3$ case is the minimal scenario without the need of 
additional symmetries.
Several new terms are added as follows:
\begin{align}
-\mathcal{L}_{Y}^{new}
&=
 \kappa_{ij} \bar e_i P_Re^c_j k^{--} + {\rm h.c.},\label{Eq:lag-yukawa3} \\
%%%%%%%%%%%%%%
V^{new}
&= \mu^2_k |k^{++}|^2 ++ \lambda_k |k^{++}|^4 + \lambda_{H k} |H|^2|k^{++}|^4
+ \lambda_{\Phi k} |\Phi_{3/2}|^2|k^{++}|^4+ \lambda_{\Phi' k} |\Phi'_{5/2}|^2|k^{++}|^4\nn\\
&
+\left[\mu (H^T\cdot \Phi_{\frac32})k^{--} +{\rm c.c.}\right]
+\left[\mu' (H^\dag \Phi_{\frac52})k^{--} +{\rm c.c.}\right],
%\label{Eq:lag-pot3}
\end{align}
where the terms $\kappa, \mu, \mu'$ mainly contribute to the appropriate 
decays into the SM particles as mentioned above.
%%%
The whole analyses have already been studied~\cite{Cheung:2017kxb}, 
and so we abbreviate this case for the moment.

\subsection{$N=5$}
In $N=5$ case, additional terms are given by
\begin{align}
-\mathcal{L}_{Y}^{new}
&=  h_{ai} \bar E_a P_R e_{i}   k^{++} + \kappa_{ij} \bar e^c_i P_R e_j k^{++} + {\rm h.c.}, \nn\\
%%%
V^{new}&= \mu^2_k |k^{++}|^2 + \mu \left[(H^\dag \Phi_{5/2})k^{++}+{\rm c.c.}\right] \nn\\
&+ \lambda_k |k^{++}|^4 + \lambda_{H k} |H|^2|k^{++}|^4
+ \lambda_{\Phi k} |\Phi_{5/2}|^2|k^{++}|^4+ \lambda_{\Phi' k} |\Phi'_{7/2}|^2|k^{++}|^4
 + {\rm h.c.},
\label{Eq:lag-yukawa5} 
\end{align} 
where the terms $h, \kappa, \mu$ mainly contribute to the 
appropriate decays into the SM particles.
Especially, $h$ can also positively contribute to the muon anomalous magnetic moment as well as LFVs.
{The formula for ${\cal M}_h$  can be expressed in a similar way as the 
generic form of ${\cal M}_f$ in Eq.~(\ref{eq:lfv-f}):}
\begin{align}
&({\cal M}_h)_{ij}
\approx
-\sum_{a=1-3}\frac{h_{aj} h^\dag_{ia} }{(4\pi)^2} 
\left[2  {F_{lfv}(M_{E_a} , m_{k} ) + 3 F_{lfv}}(m_{k},M_{E_{Ea}})\right],
\end{align} 
where $F_{lfv}$ is defined in  Eq.~(\ref{eq:lfv-lp}).

{\it Flavor-Changing Leptonic $Z$ Boson Decays}
are modified due to the contribution of $h$ and the resulting form is given by
\begin{align}
%{\color{blue}
&\text{BR}(Z\to\ell^-_i\ell^+_j)
=
\frac{G_F}{3\sqrt2 \pi} \frac{m_Z^3 s_w^4(N+1)^2}{4(16\pi^2)^2 \Gamma_{Z}^{\rm tot}} 
\left|
\sum_{a=1}^{3} f_{ia} f^*_{ja} \left[ F_2(E_a,\phi)+F_3(E_a,\phi) \right]\nn\right.\\
&\left.
+
\sum_{b=1}^{3} g_{ib} g^*_{jb} \left[ F_2(E_b,\phi')+F_3(E_b,\phi') \right]
+
\sum_{c=1}^{3} h^\dag_{ic} h^T_{jc} \left[ F_2(E_c,k^{\pm\pm})+F_3(E_c,k^{\pm\pm})
\right] 
\right|^2,
%}
\label{eq:Zll-5}
\end{align}

{\it Decay modes}:
Possible decay modes of the exotic particles are 
\begin{align}
\phi^{--}&\underbrace\to_f E^{-3} \ell^+,\quad
\phi'^{-4}\underbrace\to_g E^{-3} \ell^-,\\
%%%
E^{-3}\to 3\ell^-& : \quad (E^{-3} \underbrace{\to}_h \ell^- k^{--}), \quad (k^{--}\underbrace{\to}_\kappa \ell^-\ell^-),\\
%%%
\end{align}
where upper component of the $\Phi_{5/2}$ doublet and the lower one 
of $\Phi'_{7/2}$ mix with each other.
Both always decay into $\phi^{--}$ and/or $\phi'^{-4}$ via the 
kinetic terms, or they can directly decay into the SM leptons and 
$E^{\pm3}$ through $f$ and $g$ respectively, which are the same modes 
as the case $N=3$.

\subsection{$N=7$}
In $N=7$ case, additional terms are given by
\begin{align}
-\mathcal{L}_{Y}^{new}
&=  \kappa_{ij} \bar e^c_i P_R e_j k^{++} + {\rm h.c.}, \nn\\
%%%
V^{new}&= \mu^2_k |k^{++}|^2 + \lambda_1 \left[(H^\dag \Phi'_{9/2})k^{--}k^{--}+{\rm c.c.}\right] 
+ \lambda_2 \left[(H^T\cdot \Phi_{7/2})k^{--}k^{--}+{\rm c.c.}\right] 
\nn\\
&+ \lambda_k |k^{++}|^4 + \lambda_{H k} |H|^2|k^{++}|^4
+ \lambda_{\Phi k} |\Phi_{7/2}|^2|k^{++}|^4+ \lambda_{\Phi' k} |\Phi'_{9/2}|^2|k^{++}|^4
 + {\rm h.c.},
\label{Eq:lag-yukawa7} 
\end{align} 
where the terms $ \kappa, \lambda_1,\lambda_2$ mainly contribute to the appropriate decays into the SM particles.
%%%

{\it Decay modes}:
Possible decay modes of the exotic particles are 
\begin{align}
H^{-4}_{1/2}(Mixing\ state\ of\ \phi^{-4}_{7/2}-\phi'^{-4}_{9/2} )&\underbrace\to_{\lambda_{1/2},\ \kappa} 4\ell^- (h_{SM}),
%\quad \phi'^{-4}_{9/2} \underbrace\to_{\lambda_2} 4\ell^- (h_{SM}),\\
\end{align}
where the other modes are the same as the general $N$.

\subsection{$N=9$}
In $N=9$ case, additional terms are given by
\begin{align}
-\mathcal{L}_{Y}^{new}
&=  \kappa_{ij} \bar e^c_i P_R e_j k^{++} + {\rm h.c.}, \nn\\
%%%
V^{new}&= \mu^2_k |k^{++}|^2 + \lambda_1 \left[(H^\dag \Phi_{9/2})k^{--}k^{--}+{\rm c.c.}\right] 
\nn\\
&+ \lambda_k |k^{++}|^4 + \lambda_{H k} |H|^2|k^{++}|^4
+ \lambda_{\Phi k} |\Phi_{9/2}|^2|k^{++}|^4+ \lambda_{\Phi' k} |\Phi'_{11/2}|^2|k^{++}|^4
 + {\rm h.c.},
\label{Eq:lag-yukawa9} 
\end{align} 
where the terms $ \kappa, \lambda_1,\lambda_2$ mainly contribute to the appropriate decays into the SM particles.
%%%

{\it Decay modes}:
Possible decay modes of the exotic particles are 
\begin{align}
H^{-4}_{1/2}(Mixing\ state\ of\ \phi^{-4}_{9/2}-\phi'^{-4}_{11/2} )&\underbrace\to_{\lambda_{1},\ \kappa} 4\ell^- (h_{SM}),
%\quad \phi'^{-4}_{9/2} \underbrace\to_{\lambda_2} 4\ell^- (h_{SM}),\\
\end{align}
where the other modes are the same as the general $N$.

\section{Numerical analysis of $N=1\sim9$  }
Here we discuss the features of the global analysis for 
$N=1$ and $N=3,5,7,9$ separately, because $N=1$ involves the constraint
of the dark matter relic density.

\subsection{N=1}
The $N=1$ case is different from the other $N$'s, because of the additional
$Z_2$ symmetry imposed in order to distinguish between the Higgs field
and the $\Phi_{1/2}$. Due to the $Z_2$ symmetry, the model has a dark matter
candidate.

\begin{figure}[t]
\includegraphics[width=70mm]{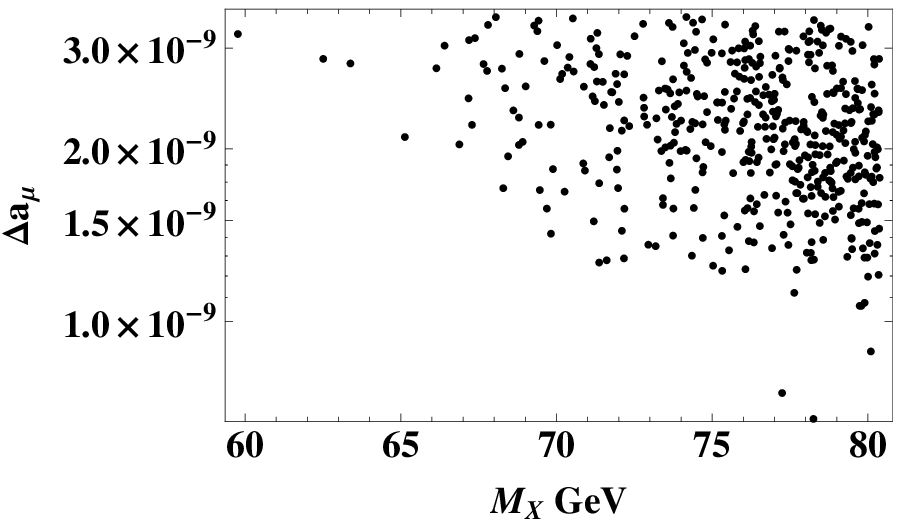}
\includegraphics[width=70mm]{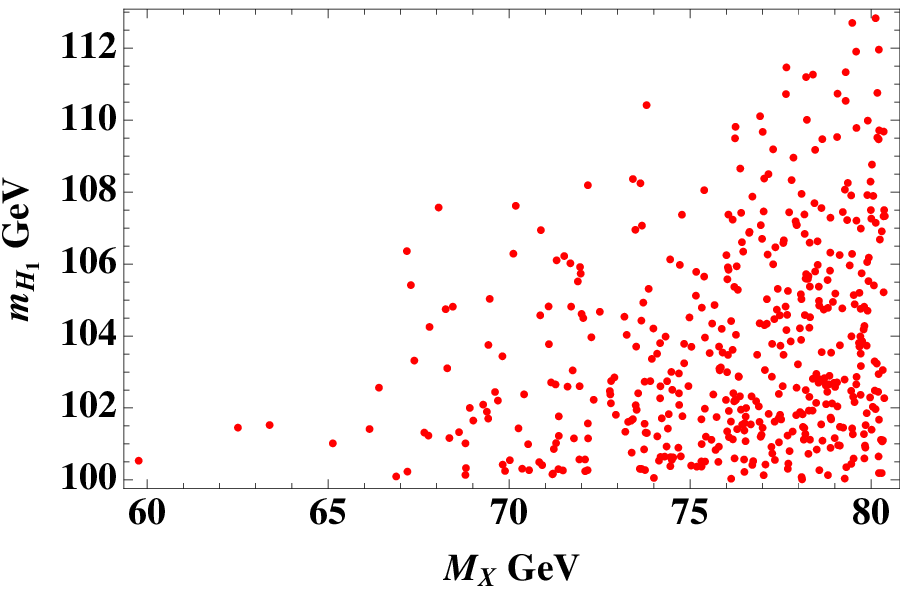}
\includegraphics[width=70mm]{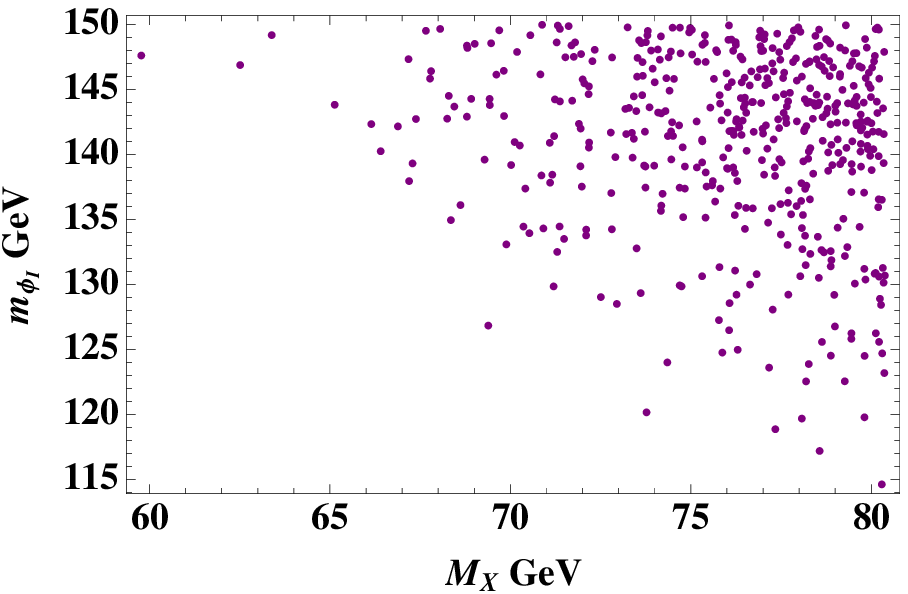}
\includegraphics[width=70mm]{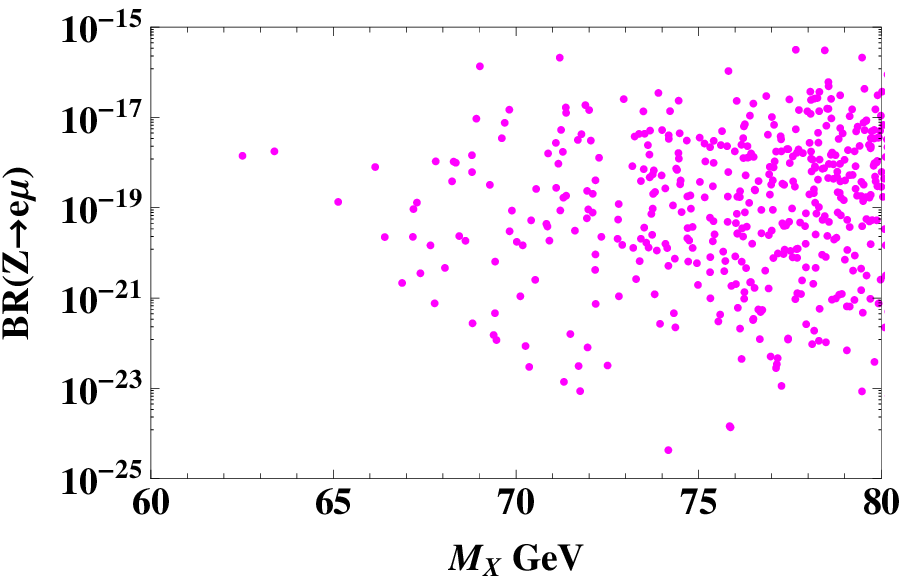}
\includegraphics[width=70mm]{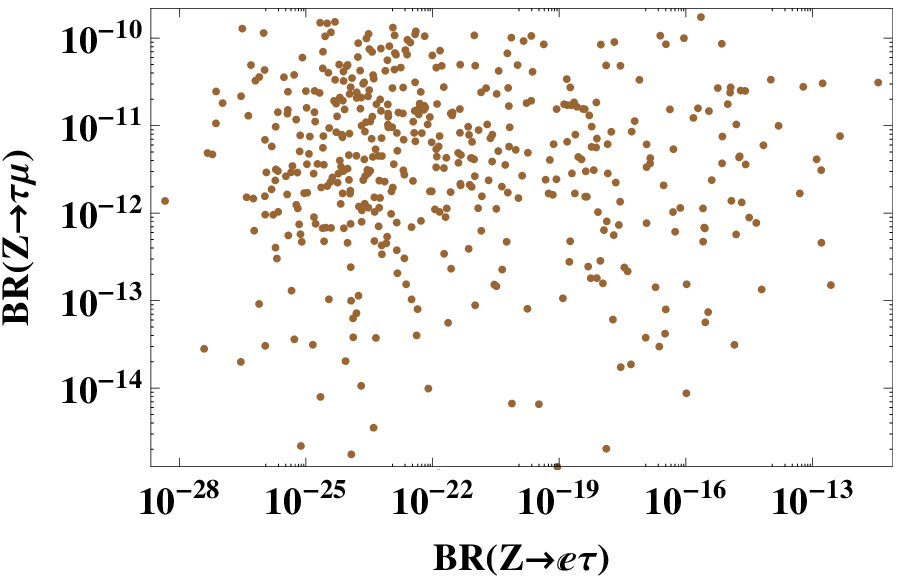}
\includegraphics[width=70mm]{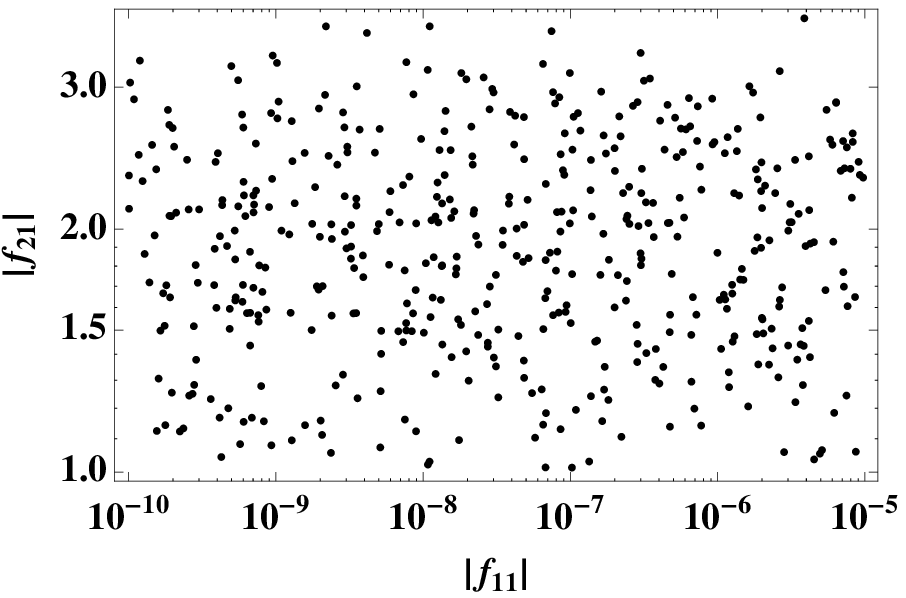}
\includegraphics[width=70mm]{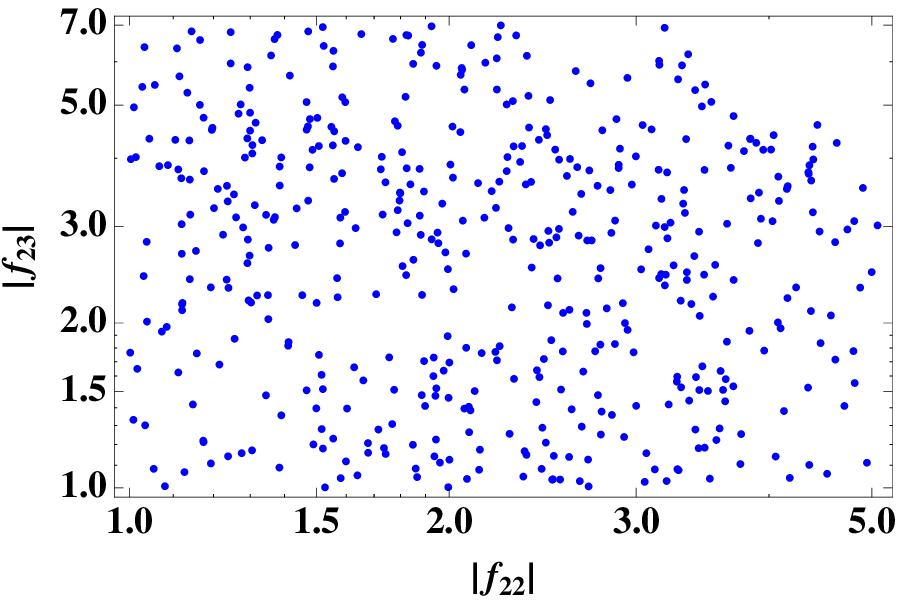}
\caption{$N=1$ case: 
Scatter plots between a pair of observables or parameters of the model,
where $0.11\le \Omega h^2\le 0.13$ is satisfied.}\label{fg:n1}
\end{figure}

The dimensionless parameters are chosen to be scanned over the 
following moderate ranges without violating naturalness or perturbativity,
as well as the mass parameters:
\begin{align}
& |s_{2\theta}|\le1,\ (|\alpha|,|\beta|,|\gamma|)\in [10^{-15},10^{-5}],\ |f_{1i}|\in [10^{-10},10^{-5}],
\ |f_{2i}|\in [1,4\pi],\ |f_{3i}|\in [10^{-3},10^{-2}], \\
 %%%
 & M_{X}\in [m_Z/2, m_W]\ {\rm GeV},\ m_{\phi_I}\in [1.2M_{X}, 150]\ {\rm GeV},\
  [m_{H_1}, m_{H_2}, m_{\phi'}]\in [100,150]\ {\rm GeV},\\
  %%%
  & M_{E_1} \in [1.2M_{X}, 150]\ {\rm GeV},
  \ M_{E_2} \in [M_{E_1}, 250]\ {\rm GeV},\ M_{E_3} \in [M_{E_2}, 500]\ {\rm GeV},
  \label{eq:out-para}
\end{align}
where we have assumed $m_{H_1}\le m_{H_2}\le  m_{\phi'}$, and $i=1\sim3$. 
We show a few plots of correlations among the observables or parameters
in Fig.~\ref{fg:n1}, where all the constraints as discussed above 
(neutrino oscillation data, LFVs, relic density with 
$0.11\le \Omega h^2\le 0.13$, {the oblique parameters,}
and $\Delta a_\mu\le(26.1+8)\times10^{-10}$ )
~\footnote{
We are content with a positive $\Delta a_\mu$ even though it is not
within the $2\sigma$ range of the data.}
are satisfied.

One remarkable feature could be found in the first panel that 
a sizable $\Delta a_\mu$ {($\Delta a_\mu > 10 \times 10^{-10}$) }
can be achieved in good agreement together with the
current relic density of DM due to the d-wave suppression,
%%%
{
where the mass of $M_X$ is between 60 and 80 GeV.
In the next panel (with the red plot), one finds that the lighter 
region of $M_X$ and the heavier mass region of $M_{H_1}$
are excluded; $m_{H_1}\lesssim$113 GeV. This could mainly arise 
from the constraint of the oblique parameters; especially $T$, 
since the $T$ parameter requires the mass degeneracy between each 
component of $\Phi_{N/2}$, and also the LFVs forbids the top-left region.
In the third panel (purple plot), the left-bottom region is also 
ruled out by the bounds on LFVs. 
In the fourth (pink plot) and fifth (brown plot) panels, all modes of 
the flavor-changing leptonic $Z$ boson decays are much below 
the current experimental bounds. This arises from the feature of 
the loop function which increases when the masses inside the loop are heavier.
This result can be important in the other cases of $N$.
In the sixth (black plot) and seventh (blue plot) panels, 
$|f_{2i}|(i=1-3)$ are restricted to be
$f_{21}\lesssim\sqrt{4\pi}$, $f_{22}\lesssim5$, and 
$f_{23}\lesssim7$. These bounds originate from LFVs.
Note here that all the other parameters run all over 
the range that we have taken in Eq.~(\ref{eq:out-para}). 
}
%%%

%{The mass of $M_X$ is between \textcolor{blue}{60} and 80 GeV while the $\phi_I$ is between 120 and 150 GeV. The lepton-changing leptonic $Z$ decays are many orders of magnitude below the current limits, with the largest one BR$(Z \to \tau \mu) \alt 10^{-10}$. Among the Yukawa couplings $f_{21,22,23} \sim O(1)$ are much larger  than the others so as to obtain sizable $\Delta a_\mu$.}

\subsection{$N=3,5,7,9$}
\begin{figure}[t]
\includegraphics[width=70mm]{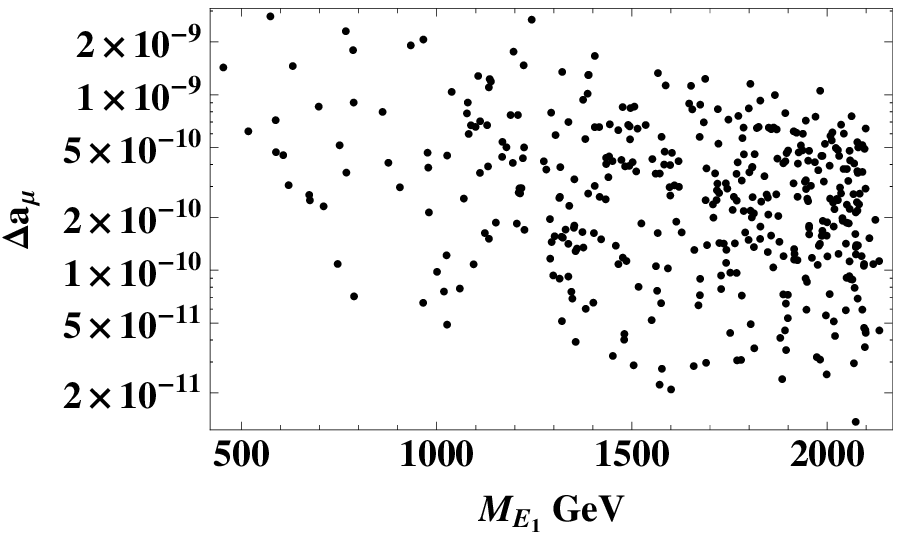}
\includegraphics[width=70mm]{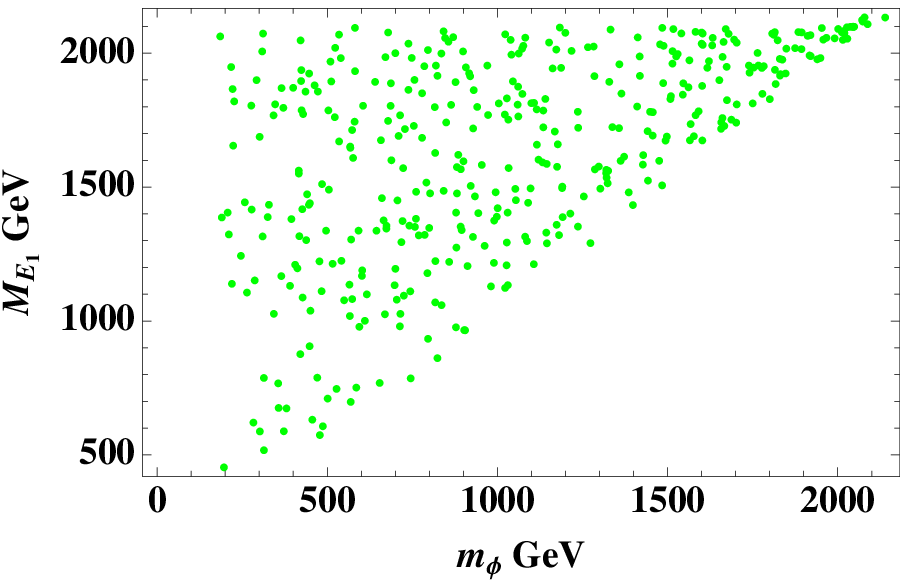}
\includegraphics[width=70mm]{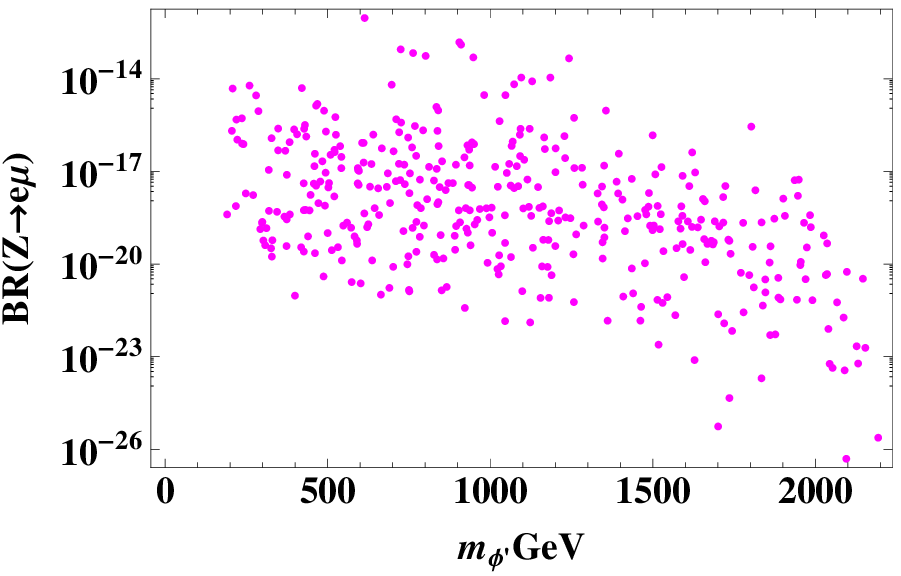}
\includegraphics[width=70mm]{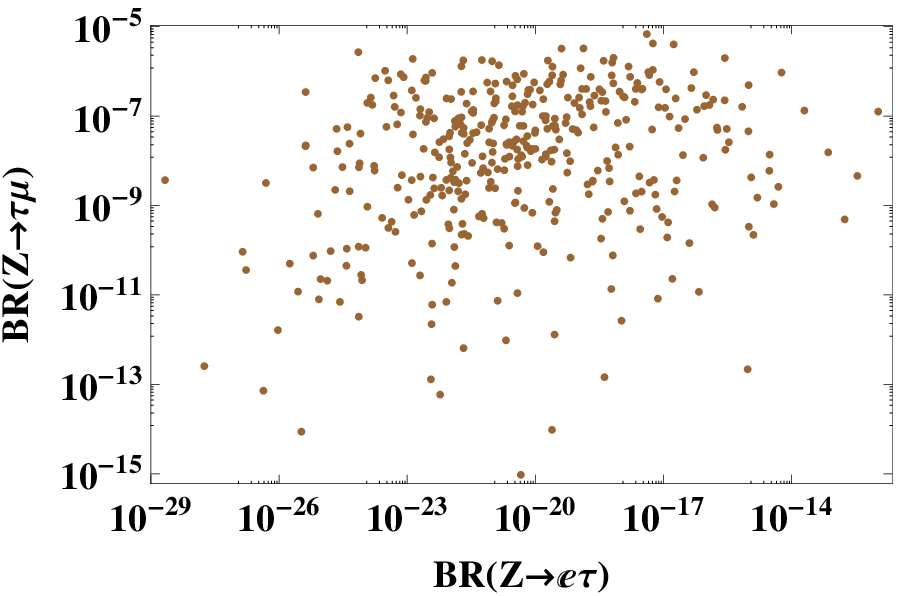}
\caption{$N=3$ case: 
Scatter plots between a pair of observables or parameters of the model.
}\label{fg:n3}
\end{figure}

\begin{figure}[t]
\includegraphics[width=70mm]{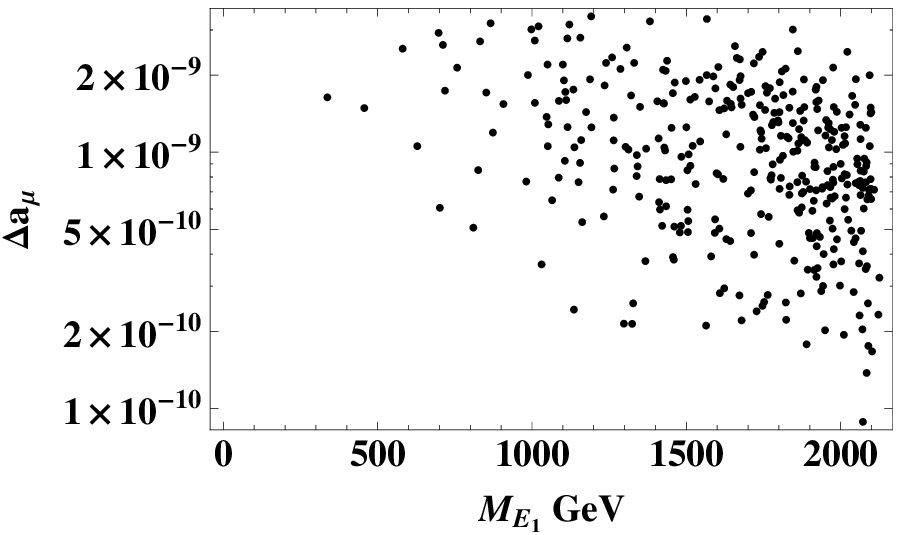}
\includegraphics[width=70mm]{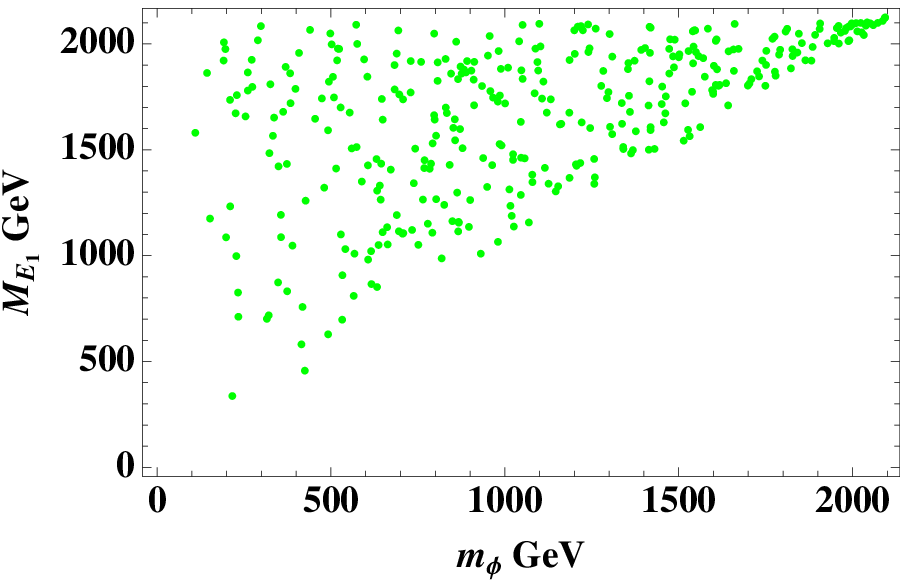}
\includegraphics[width=70mm]{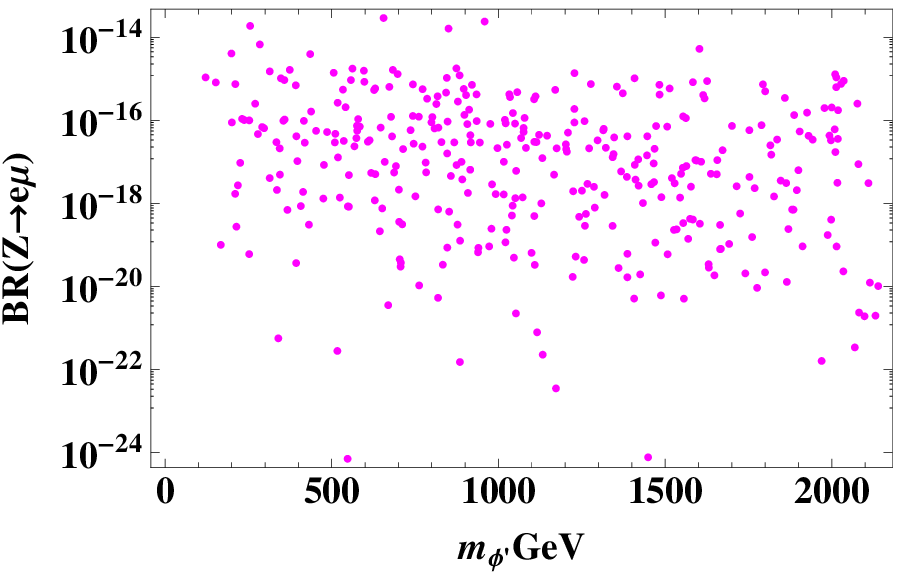}
\includegraphics[width=70mm]{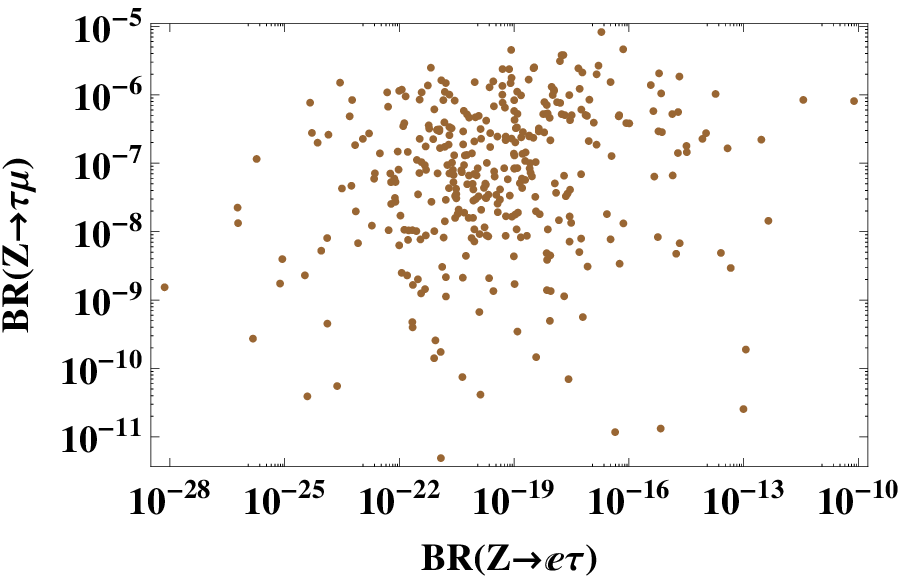}
\caption{$N=5$ case: 
Scatter plots between a pair of observables or parameters of the model.
}\label{fg:n5}
\end{figure}

\begin{figure}[t]
\includegraphics[width=70mm]{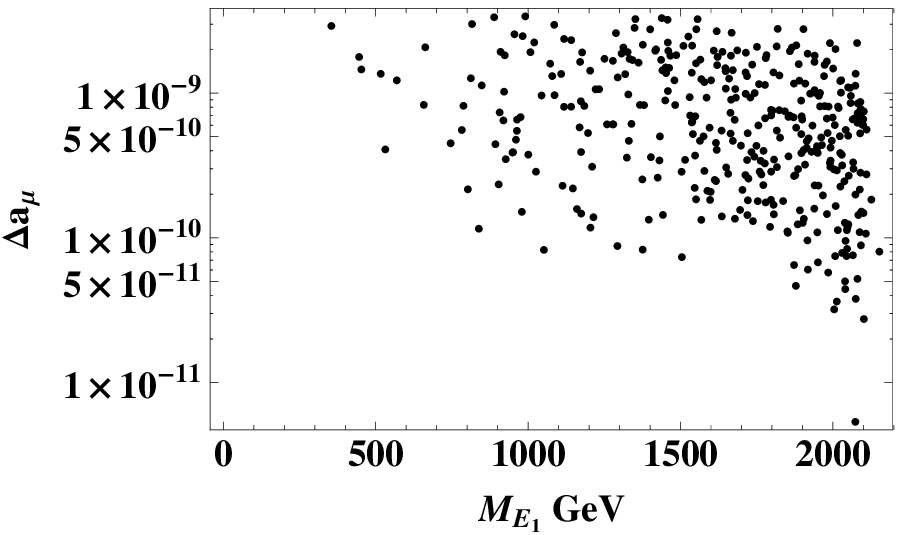}
\includegraphics[width=70mm]{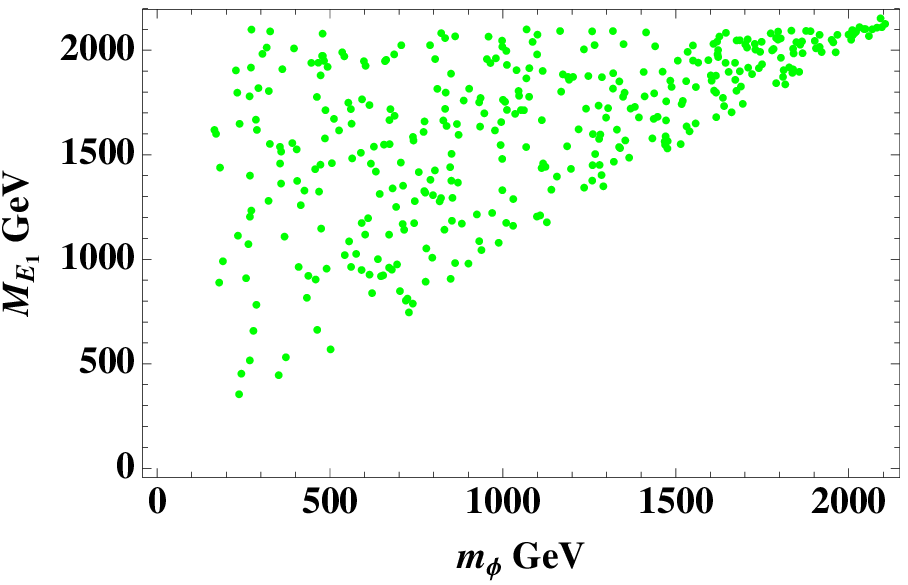}
\includegraphics[width=70mm]{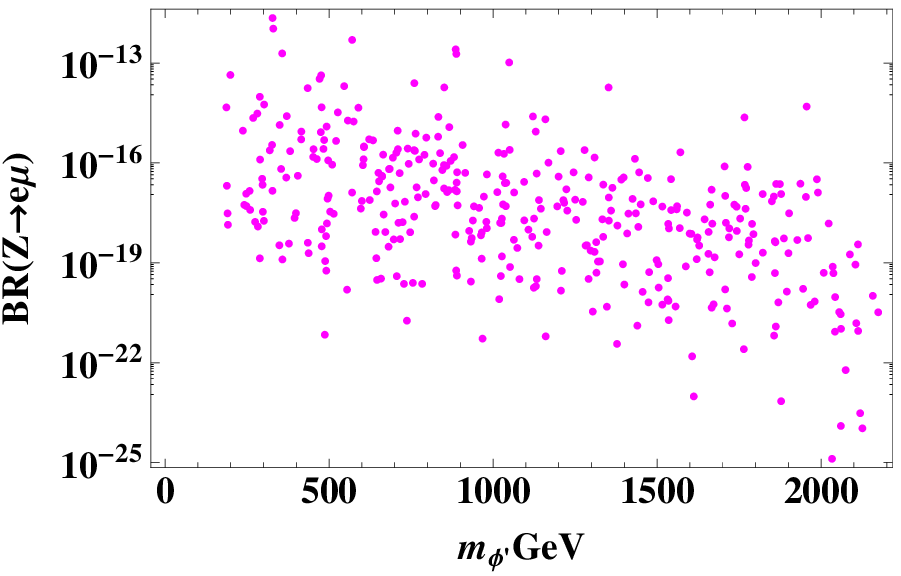}
\includegraphics[width=70mm]{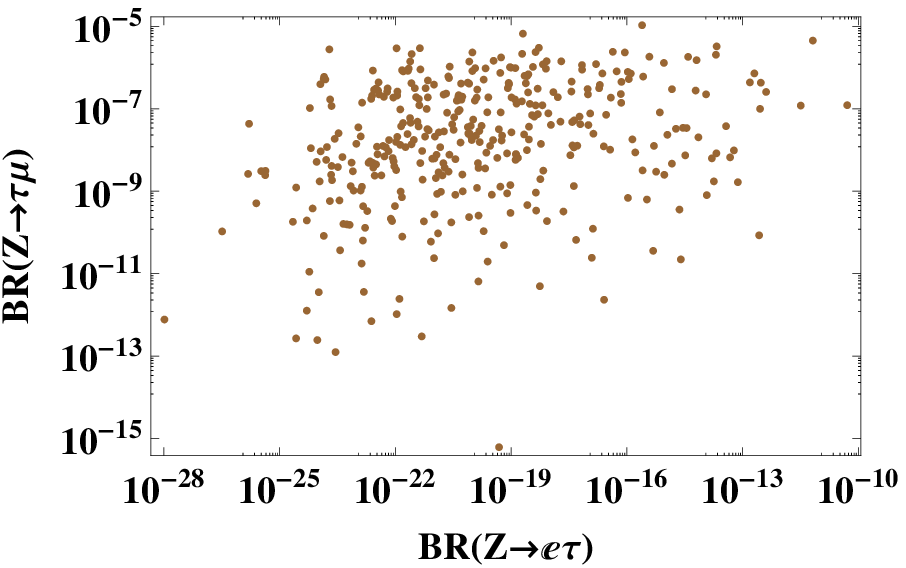}
\caption{$N=7$ case: 
Scatter plots between a pair of observables or parameters of the model.
}\label{fg:n7}
\end{figure}

\begin{figure}[t]
\includegraphics[width=70mm]{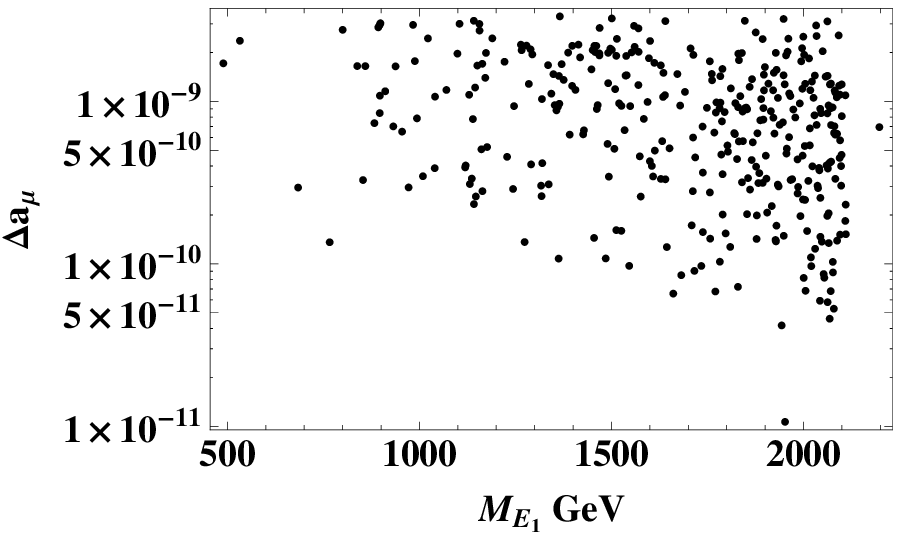}
\includegraphics[width=70mm]{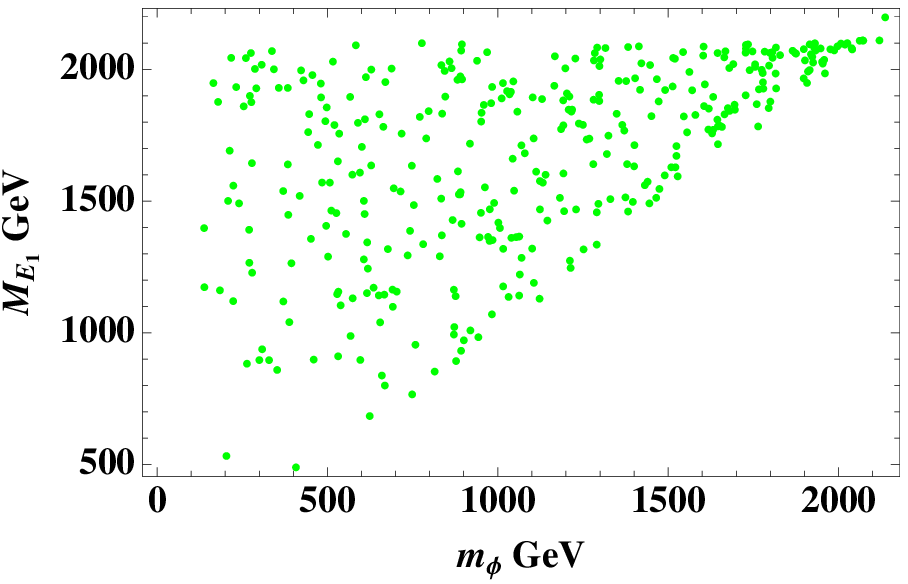}
\includegraphics[width=70mm]{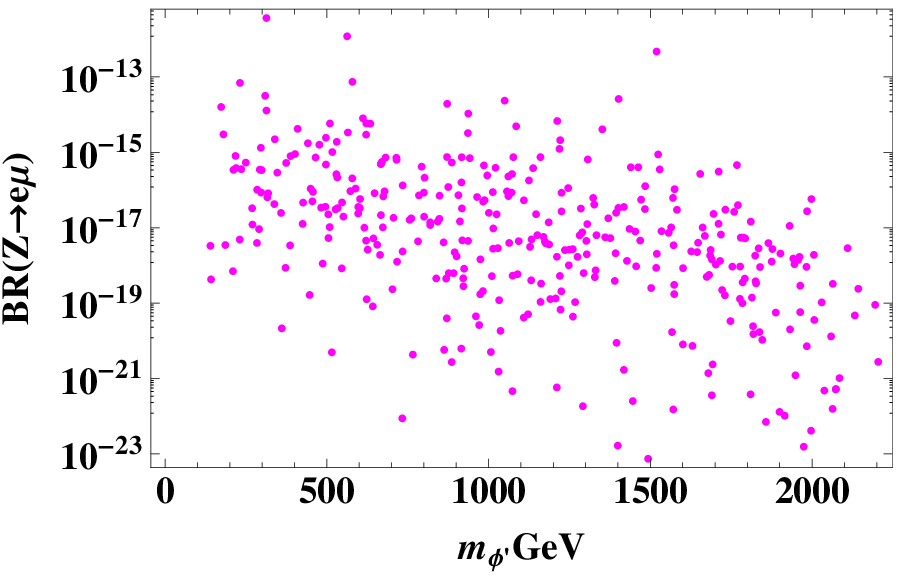}
\includegraphics[width=70mm]{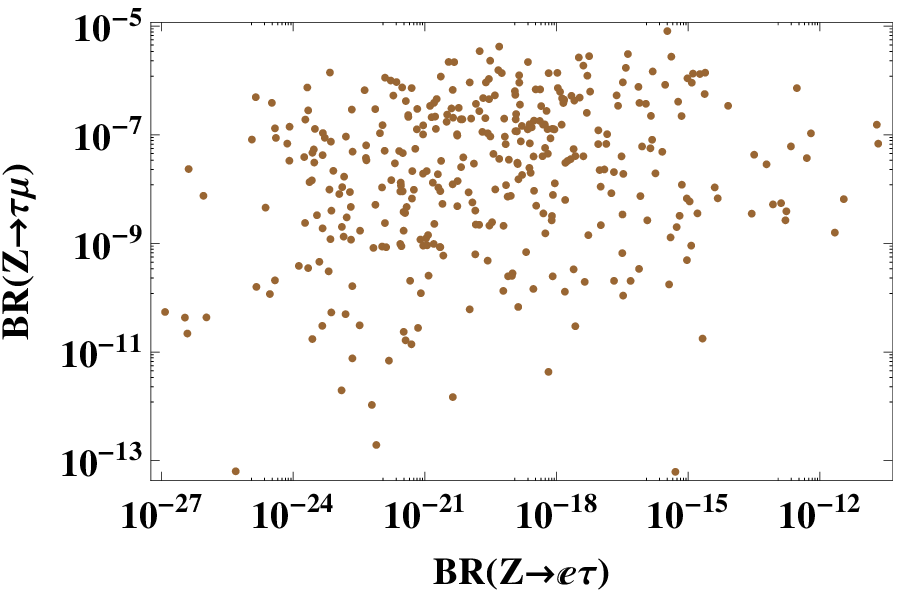}
\caption{$N=9$ case: 
Scatter plots between a pair of observables or parameters of the model.
}\label{fg:n9}
\end{figure}

Next we investigate the case of $N=3,5,7,9$.
The dimensionless parameters are scanned similarly to the $N=1$ case, and
the mass parameters as well:
\begin{align}
%|s_{2\theta}|\le1,\quad (|\alpha|,|\beta|,|\gamma|)\in [10^{-15},10^{-5}],\quad
 &|f(h)_{1i}|\in [10^{-10},10^{-5}],\quad |f(h)_{2i}|\in [1,4\pi],\quad
 |f(h)_{3i}|\in [10^{-3},1],\\
 %%%
 & m_{H_1}\in [100,2000]\ {\rm GeV},\quad m_k \in [500,2000]\ {\rm GeV}
 ,\quad M_{E_a} \in [m_{\phi'},2100]\ {\rm GeV}, \label{eq:output_N3-9}
\end{align}
where we have assumed $m_{H_1}\le m_{H_2}\le m_{\phi}\le m_{\phi'}$, 
$M_{E_1}\le M_{E_2}\le M_{E_3}$,
 $i=1\sim3$ and the others are the same as the case of $N=1$.
Under these ranges, we show a number of plots in 
Figs.~\ref{fg:n3} -- \ref{fg:n9} for $N=3 - 9$, respectively.
Several features are as follows:
 %%%%%%%%%%%

\begin{enumerate}
\item
In the first panel of all Figs.~\ref{fg:n3} -- \ref{fg:n9},
the majority of the allowed points can achieve positive $\Delta a_\mu$ larger 
than $1\times 10^{-10}$ for $N=3-9$, {and the 
allowed region that satisfies  $\Delta a_\mu$
is wider when the number of $N$ increases.}

\item In the upper-right panel of all Figs.~\ref{fg:n3} -- \ref{fg:n9}
shows that $M_{E_1} > M_{\phi}$, {whose mass hierarchy provides the larger value of loop function in the LFVs}, because of the requirement of 
positive $\Delta a_\mu$, as indicated in Eq.~(\ref{eq:lfv-f}).

\item In the bottom panels of all Figs.~\ref{fg:n3} -- \ref{fg:n9},
the largest flavor-changing leptonic $Z$ decays is $Z \to \tau\mu$,
which can be as large as $10^{-5}$, while the other two decay
branching ratios are many orders of magnitude below the current limits.
The expected value for BR$(Z \to \tau\mu)$ almost reaches the current 
experimental bound in Eq.(\ref{eq:zmt-exp}) that could be tested by 
a Giga-Z type experiment at lepton colliders~\cite{AguilarSaavedra:2001rg}.
{The large value mainly arises from the heavier masses 
inside the loop functions in addition to the larger Yukawa couplings. 
This is one of the important differences between $N=1$ and the other $N$'s.}

\item 
We observe that ${\rm BR}(Z\to\tau^\pm\mu^\mp)$ increases
with increase in $N$, which indeed links to larger $\Delta a_\mu$, 
with larger $|f_{2i}|$.
Similarly, larger ${\rm BR}(Z\to e^\pm\mu^\mp)$ and 
${\rm BR}(Z\to\tau^\pm e^\mp)$ are obtained as $N$ increases.
However, they are far below the current experimental bounds in Eq.(\ref{eq:zmt-exp}).

%\item Among the Yukawa couplings, $f_{21,22,23}$ are the largest and of order $O(1-10)$ because of the requirement of $\Delta a_\mu$, while the others are very small due to lepton-flavor violation constraints.

\item With increasing $N$, the number of allowed parameter-space points 
decreases. {It mainly arises from the constraints of LFVs 
and the oblique parameters.}

%\item 
%Note that the allowed mass region between $m_{H_1}-m_{H_2}$ is restricted by the $T$-parameter such that they are close to each other.

\item For $N=5$ there are additional interactions
among $k^{++}$, $E_a$ and $e_i$, which contribute to the lepton-flavor
violation and $\Delta a_\mu$ in the same way as the $f$ terms in Eq.~(\ref{eq:lfv-f}).
Therefore, the behavior of the plots for $h_{ij}$ would be very similar
to those of $f_{ij}$,

 \end{enumerate}
{
All the other parameters run all over the ranges that we have taken 
in Eq.(\ref{eq:output_N3-9}), and the allowed mass region 
between $m_{H_1}-m_{H_2}$ is restricted 
by the $T$-parameter such that they are close to each other.}

 %%%%%%%%%%%

%%%%%%%
%%%%%%%
%%%%%%%
%%%%%%%

\section{Collider Signals}
We expect that the largest production rate of the exotic particles comes 
from pair production of the lightest fermions $E_1$.
Drell-Yan (DY) production of $E_1 \overline{E}_1$ occurs 
via $\gamma,\, Z$ exchanges. The interactions can be obtained from the
kinetic term of the fermion $E_1$. Since $E_1$ is a singlet, the interactions
with $\gamma$ and $Z$ are given by
\[
  {\cal L}  = - e \overline{E}_1 \gamma^\mu Q_E E_1\, A_\mu 
              + \frac{g s_W^2 }{c_W} \overline{E}_1 \gamma^\mu Q_E E_1
  \, Z_\mu \;,
\]
where $s_W$ and $c_W$ are respectively the sine and cosine of the Weinberg
angle, and $Q_E$ is the electric charge of the fermion $E_1$. The following
applies for $N=1,3,5,7,9$ and $E = E_1$ for simplicity.

The square of the scattering amplitude, summed over spins, for 
$q (p_1) \bar q (p_2) \to E (k_1) \overline{E} (k_2)$ can be written as
\begin{eqnarray}
\sum |{\cal M}|^2 &=& 4 e^4 Q_E^2 \left[ 
   \left(\hat u - M^2_E \right )^2 +    \left(\hat t - M^2_E \right )^2 
  + 2 \hat s M_E^2 \right ] \nonumber \\
& \times & \left\{
        \left| \frac{Q_q}{\hat s} - \frac{g_L^q}{ c_W^2 } \frac{1}{\hat s -m_Z^2}
             \right|^2
+
        \left| \frac{Q_q}{\hat s} - \frac{g_R^q}{ c_W^2 } \frac{1}{\hat s -m_Z^2}
             \right|^2 
  \right \} \;,
\end{eqnarray}
where $\hat s, \, \hat t, \, \hat u$ are the usual Mandelstam variables
for the subprocess, and 
$g_{L,R}^q$ are the chiral couplings of quarks to the $Z$ boson.
The subprocess differential cross section is given by
\begin{eqnarray}
\frac{d \hat \sigma}{ d\cos \hat \theta} &=& \frac{\beta  e^4 Q_E^2}{96\pi}
 \left[ 
   \left(\hat u - M^2_E \right )^2 +    \left(\hat t - M^2_E \right )^2 
  + 2 \hat s M_E^2 \right ] \nonumber \\
& \times & \left\{
      \left| \frac{Q_q}{\hat s} - \frac{g_L^q}{ c_W^2 } \frac{1}{\hat s -m_Z^2}
             \right|^2
+
        \left| \frac{Q_q}{\hat s} - \frac{g_R^q}{ c_W^2 } \frac{1}{\hat s -m_Z^2}
             \right|^2 
  \right \} \;,
\end{eqnarray}
where $\beta = \sqrt{ 1- 4 M_E^2 / \hat s}$,
and $g_{L}^q = T_{3q} - s_W^2 Q_q$, $g_{R}^q = - s_W^2 Q_q$,
where $T_{3q}$ is the third component of the isospin of $q$.
This subprocess cross section is then folded with parton distribution 
functions to obtain the scattering cross section at the $pp$ collision level.
The $K$ factor for the production cross sections is expected to be similar
to the conventional DY process, which is approximately $K \simeq 1.3$ at 
the LHC energies.
The production cross sections for 
$pp \to E_1 \bar E_1$ at $\sqrt{s}=13$ TeV scale as $Q_E^2$, where 
$Q_E = (N+1)/2$.  In Ref.~\cite{Cheung:2017kxb}, the production cross
sections for $E^{--} E^{++}$ are shown, and so therefore for other $N=1,5,7,9$
can be easily estimated by simple scaling.

We can also derive the decay width of $E_1$ for a general $N$. The major decay
modes are those via the terms $f_{i1}$ and $g_{i1}$ in the Lagrangian
in Eq.~(\ref{Eq:lag-yukawa}). The decay modes are 
\[
E^{-(N+1)/2}_1 \to \nu_i \phi^{-(N+1)/2}_{N/2},\; \ell^-_i \phi^{-(N-1)/2}_{N/2} ,\;
 \nu_i {\phi'}_{N'/2}^{-(N+1)/2},\; \ell^+_i {\phi'}_{N'/2}^{-(N+3)/2}.
\]
 The decay width is given by
\begin{equation}
\label{widthE}
\Gamma(E_1) = \frac{M_{E_1}}{32\pi}  \left \{
\left(  1 - \frac{ m^2_{\phi} }{ M^2_{E_1}} \right ) \, 
     \sum_{i=1}^{3}  |f_{i1}|^2 
+ 
\left(  1 - \frac{m^2_{\phi^{'}} }{ M^2_{E_1}} \right ) \, 
     \sum_{i=1}^3  |g_{i1}|^2 
\right \}.
\end{equation}
Nevertheless, the contributions from $f_{i1}$ dominate because of the
requirement of the $\Delta a_\mu$.

Here we start with the detection of the final states for each
case of $N=1,3,5,7,9$.  We focus on the lightest state $E_1$.

\subsubsection{$N=1$}
The exotic scalar doublet fields take the form:
\[
   \phi_{1/2} = \left(  \begin{array}{cc}
                             \phi^+_{1/2} \\
                             \phi^0_{1/2} \\
                        \end{array}  \right ) , \qquad
   \phi'_{3/2} = \left(  \begin{array}{cc}
                             \phi^{'++}_{3/2} \\
                             \phi^{'+}_{3/2} \\
                        \end{array}  \right ) .
\]
The neutral component $\phi^0_{1/2}$ is stable because of the $Z_2$ symmetry.
The exotic fermion $E^-$ can decay via the terms $f_{i1}$ and $g_{i1}$:
\begin{equation}
 - {\cal L} \supset 
  f_{i1} \left( \bar \nu_i \phi^+_{1/2} + \bar \ell_i \phi^0_{1/2} 
          \right) \, E_{1_R}
+ g_{i1} \left( \bar \nu_i \phi^{'-}_{3/2} - \bar \ell_i \phi^{'--}_{3/2} 
        \right) \, E_{1_R}^c 
+ {H.c.}.
\end{equation}
Assuming $|f| \gg |g|$ due to the requirement of $\Delta a_\mu$, 
the exotic fermion $E^-_1$ decays via the term $f_{i1}$ as 
 \[
   E^-_1 \to \nu \phi^-_{1/2} ,\;\;\;  \ell^- \phi^0_{1/2}  \;,
\]
where $\phi^-_{1/2} \to \phi^0_{1/2} W^*$ and $\phi^0_{1/2}$ is the lightest
exotic particle being stable. Therefore, Drell-Yan production of
$E^-_1 E^+_1$ gives rise to a pair of charged leptons (not necessary of the 
same flavor) plus missing energies.

\subsubsection{$N=3$}
This particular case has been given in detail in Ref.~\cite{Cheung:2017kxb}.
In summary, the exotic fermion pair $E^{--}_1 E^{++}_1$ is produced via the
Drell-Yan process.  The fermion $E^{--}_1$ decays via the terms $f_{i1}$ 
and $g_{i1}$:
\begin{equation}
 - {\cal L} \supset 
  f_{i1} \left( \bar \nu_i \phi^{++}_{3/2} + \bar \ell_i \phi^+_{3/2} 
          \right) \, E^{--}_{1_R}
+ g_{i1} \left( \bar \nu_i {\phi'}^{--}_{5/2} - \bar \ell_i {\phi'}^{3-}_{5/2} 
        \right) \, E_{1_R}^c 
+ {H.c.}.
\end{equation}
The decays via the dominant $f_{i1}$ term are
\[
  E^{--}_1 \to \ell^- \phi_{3/2}^-,\;\;\; \nu \phi_{3/2}^{--},
\]
followed by (via the mixing with scalar $k^{--}$)
\[
  \phi_{3/2}^{--} \to \ell^- \ell^- 
\]
\[
  \phi_{3/2}^{-} \to \phi_{3/2}^{--} W^+ \to \ell^- \ell^- W^+.
\]
Each $E^{--}_1$ decays into two or four charged leptons. Therefore, the
final state can consist of 4, 6, or 8 charged leptons plus missing energies.

\subsubsection{$N=5$}
This case is similar to $N=3$. The relevant terms in the Lagrangian 
responsible for the decay of the exotic fermion $E^{3-}$ are
\begin{equation}
 - {\cal L} \supset 
  f_{i1} \left( \bar \nu_i \phi^{3+}_{5/2} + \bar \ell_i \phi^{2+}_{5/2} 
          \right) \, E^{3-}_{1_R}
+ g_{i1} \left( \bar \nu_i {\phi'}^{3-}_{7/2} - \bar \ell_i {\phi'}^{4-}_{7/2} 
        \right) \, E_{1_R}^c 
+ h_{1i} \bar E_1 P_R \ell_i k^{--} + \kappa_{ij} \bar {e}^c_i P_R e_j k^{++}
+ {H.c.} ,
\end{equation}
in which $\phi_{5/2}^{2+}$ and $k^{++}$ mix. The exotic fermion $E^{3-}_1$ so
produced will decay via
\[
 E^{3-}_1 \to \nu \phi_{5/2}^{3-} \to \nu \phi_{5/2}^{2-} W^- \to 
  \nu\ell^- \ell^- W^-
\]
\[
E^{3-}_1 \to \ell^- \phi^{2-},\, \ell^- k^{--} \to \ell^- \ell^- \ell^-.
\]
Each $E^{3-}_1$ decays into 3 charged leptons, without or with small missing
energy. Therefore, the final state of $E^{3-} E^{3+}$ pair production
consists of 6 charged leptons, mainly without missing energies.

\subsubsection{$N=7$}
In this case, there are additional terms in the Lagrangian that allow
the quartic-charged $\phi$ or $\phi'$ to decay into a pair of $k^{++} k^{++}$,
which further decay into 4 charged leptons.
\begin{eqnarray}
 - {\cal L} & \supset &
  f_{i1} \left( \bar \nu_i \phi^{4+}_{7/2} + \bar \ell_i \phi^{3+}_{7/2} 
          \right) \, E^{4-}_{1_R}
+ g_{i1} \left( \bar \nu_i {\phi'}^{4-}_{9/2} - \bar \ell_i {\phi'}^{5-}_{9/2} 
        \right) \, E_{1_R}^c  \nonumber \\
&& 
+ \lambda_1 v {\phi'}^{4+} k^{--} k^{--} 
+ \lambda_2 v {\phi}^{4+} k^{--} k^{--} 
+ \kappa_{ij} \bar {e}^c_i P_R e_j k^{++}
+ {H.c.} .
\end{eqnarray}
Each exotic fermion $E^{4-}_1$ decays via
\[
  E^{4-}_1 \to \nu \phi_{7/2}^{4-} \to \nu k^{--} k^{--} \to 
   \nu \ell^- \ell^- \ell^- \ell^- 
\]
\[
  E^{4-}_1 \to \ell^- \phi_{7/2}^{3-} \to \ell^- \phi_{7/2}^{4-} W^+ 
\to \ell^- k^{--} k^{--} W^+ \to \ell^- \ell^- \ell^- \ell^- \ell^- W^+.
\]
Thus, each $E^{4-}_1$ can decay into 4 or 6 charged leptons. Therefore, the 
final state of $E^{4-}_1 E^{4+}_1$ production consists of 8, 10, or 12
charged leptons.

\subsubsection{$N=9$}
In this case, there is one additional term in the Lagrangian that allows
the quartic-charged $\phi$ to decay into a pair of $k^{++} k^{++}$,
which further decay into 4 charged leptons.
\begin{equation}
 - {\cal L} \supset 
  f_{i1} \left( \bar \nu_i \phi^{5+}_{9/2} + \bar \ell_i \phi^{4+}_{9/2} 
          \right) \, E^{5-}_{1_R}
+ g_{i1} \left( \bar \nu_i {\phi'}^{5-}_{11/2} - \bar \ell_i {\phi'}^{6-}_{11/2} 
        \right) \, E_{1_R}^c 
+ \lambda_1 v {\phi}^{4+}_{9/2} k^{--} k^{--} 
+ \kappa_{ij} \bar {e}^c_i P_R e_j k^{++}
+ {H.c.} .
\end{equation}
Each exotic fermion $E^{5-}_1$ decays via
\[
  E^{5-}_1 \to \nu \phi_{9/2}^{5-} \to \nu \phi_{9/2}^{4-} W^-  \to 
\nu k^{--} k^{--} W^- \to \nu \ell^- \ell^- \ell^- \ell^-  W^-
\]
\[
 E^{5-}_1 \to \ell^- \phi_{9/2}^{4-} \to \ell^- \ell^- \ell^- \ell^- \ell^-.
\]
Thus, each $E^{5-}$ decays into 5 charged leptons. Therefore, the 
final state of $E^{5-}_1 E^{5+}_1$ production consists of 10 charged leptons.

\section{Conclusions}

We have generalized a radiative neutrino-mass model by introducing 3
generations of fermion pairs $E^{-(N+1)/2} E^{+(N+1)/2}$ and a couple
of multi-charged bosonic doublet fields $\Phi_{N/2}, \Phi_{N/2+1}$,
where $N=1,3,5,7,9$. We have shown that the model can satisfy the
neutrino masses and oscillations, lepton-flavor violations, the
oblique parameters, and the $\Delta a_\mu$.  We also made predictions for
the collider signatures of the model.  In general, searches for multi
charged leptons in the final state, not necessarily of the same
flavor, are interesting probes of the model.

We offer a few more comments as follows.
\begin{enumerate}

\item For $N=1$, where there is an additional $Z_2$ symmetry to distinguish
the Higgs field from $\Phi_{1/2}$, there is a possible dark matter candidate.
On the other hand, for $N=3,5,7,9$ cases we require extra terms in the
Lagrangian in order to make sure no stable charged particles remain.
The real part of the neutral component of $\phi_1$ serves as the dark matter,
and its mass is around ${60}-80$ GeV, in order to satisfy the relic density
of the Universe.

\item In order to give large positive contributions to the $\Delta a_\mu$ 
the relevant components in $f$, such as $f_{21}$, $f_{22}$ 
and $f_{23}$, have to be large enough.  On the other hand, other
Yukawa couplings are in general very small because of lepton-flavor
violation constraints.

\item The largest flavor-changing leptonic $Z$ decays is 
$Z \to \tau \mu$, which can have a branching ratio  as large as $10^{-5}$.
It can be tested in the Giga-Z option in the future $e^+ e^-$ colliders.

\item The mass splitting among the components in $\Phi$ doublet fields
is restricted because of the $T$ parameters.

\item Drell-Yan production of exotic fermion pairs would give rise
to multi charged leptons in the final state.  In general, a larger $N$ would
give more charged leptons in the final state.

\end{enumerate}

\section*{Acknowledgments}
We thank Chang-Hun Lee for discussion.
This work was supported by the Ministry of Science and Technology
of Taiwan under Grants No. MOST-105-2112-M-007-028-MY3.

%\section*{ Appendix}
%%%%%%%%%%%%%%%%%%%...

\end{document}